\documentclass[preprint,10pt,time]{elsarticle}

\usepackage[twoside,
  paperwidth=210mm,
  paperheight=297mm,
  textheight=622pt,
  textwidth=468pt,
  centering,
  headheight=50pt,
  headsep=12pt,
  footskip=18pt,
  footnotesep=24pt plus 2pt minus 12pt,
  columnsep=2pc,
  marginparwidth=40pt]{geometry}

\journal{Journal of the Mechanics and Physics of Solids}

\usepackage{multirow}
\usepackage{hyperref}
\usepackage{dblfloatfix}
\usepackage[utf8]{inputenc}

\usepackage{amssymb}
\usepackage{amsmath}
\usepackage{upgreek}
\usepackage{algorithm}
\usepackage{array}
\usepackage{bbold}

\usepackage{graphicx}
\usepackage{caption}
\usepackage{subcaption}
\usepackage[usenames]{color}
\usepackage[labelfont=bf,textfont=normalsize]{caption}
\usepackage[labelfont=bf,textfont=small]{subcaption}

\usepackage{here}
\usepackage{multicol}
\usepackage{wrapfig}

\usepackage{soul}
\usepackage{comment}
\usepackage{booktabs}

\usepackage{color}
\usepackage[normalem]{ulem}
\newcommand{\Delete} [1]{\bgroup\noindent\textcolor{red}{\xout{#1}}\egroup\ignorespacesafterend}
\newcommand{\Insert} [1]{\bgroup\noindent\textcolor{blue}{#1}\egroup\ignorespacesafterend}

\date{2021}

\begin{document}

\begin{frontmatter}

%% Title, authors and addresses

%% use the tnoteref command within \title for footnotes;
%% use the tnotetext command for theassociated footnote;
%% use the fnref command within \author or \address for footnotes;
%% use the fntext command for theassociated footnote;
%% use the corref command within \author for corresponding author footnotes;
%% use the cortext command for theassociated footnote;
%% use the ead command for the email address,
%% and the form \ead[url] for the home page:
%% \title{Title\tnoteref{label1}}
%% \tnotetext[label1]{}
%% \author{Name\corref{cor1}\fnref{label2}}
%% \ead{email address}
%% \ead[url]{home page}
%% \fntext[label2]{}
%% \cortext[cor1]{}
%% \address{Address\fnref{label3}}
%% \fntext[label3]{}

\title{Identification of dislocation reaction kinetics in complex dislocation networks for continuum modeling using data-driven methods}

%% use optional labels to link authors explicitly to addresses:
%% \author[label1,label2]{}
%% \address[label1]{}
%% \address[label2]{}

\cortext[cor1]{Corresponding author.}

\author[IAM]{Balduin Katzer}
\author[IAM]{Kolja Zoller}
\author[IAM]{Daniel Weygand}
\author[IAM,HS]{Katrin Schulz \corref{cor1}}

\address[IAM]{Karlsruhe Institute of Technology (KIT), Institute for Applied Materials (IAM),\\
              Kaiserstr. 12, 
              76131 Karlsruhe, Germany}
\address[HS]{Karlsruhe University of Applied Sciences, Moltkestr. 30, 76133, Karlsruhe, Germany}
\ead{katrin.schulz@kit.edu}

%\author[institute]{name}
%\address[institute]{name of institute,
%              Karlsruhe Institute of Technology (KIT),\\
%              Kaiserstr. 12, 
%              76131 Karlsruhe, Germany}

\begin{abstract}
%%% Text of abstract

Plastic deformation of metals involves the complex evolution of dislocations forming strongly connected dislocation networks.
These dislocation networks are based on dislocation reactions, which can form junctions during the interactions of different slip systems. 
Extracting the fundamentals of the network behaviour during plastic deformation by adequate physically based theories is essential for crystal plasticity models. 
In this work, we demonstrate how knowledge from discrete dislocation dynamics simulations to continuum-based formulations can be transferred by applying a physically based dislocation network evolution theory. 
By using data-driven methods, we validate a slip system dependent rate formulation of network evolution.
We analyze different discrete dislocation dynamics simulation data sets of face-centred cubic single-crystals in high symmetric and non-high symmetric orientations under uniaxial tensile loading. 
Here, we focus on the reaction evolution during stage II plastic deformation.
Our physically based model for network evolution depends on the plastic shear rate and the dislocation travel distance described by the dislocation density. 
We reveal a dependence of the reaction kinetics on the crystal orientation and the activity of the interacting slip systems, which can be described by the Schmid factor.
It has been found, that the generation of new reaction density is mainly driven by active slip systems. 
However, the deposition of generated reaction density is not necessarily dependent on the slip system activity of the considered slip system, i.e. we observe a deposition of reaction density on inactive slip systems especially for glissile and coplanar reactions.

%CORRECTION DONE

\end{abstract}

%\begin{keyword}
%%% keywords here, in the form: keyword \sep keyword
%%% PACS codes here, in the form: \PACS code \sep code
%%% MSC codes here, in the form: \MSC code \sep code
%%% or \MSC[2008] code \sep code (2000 is the default)
%\end{keyword}

\end{frontmatter}

%% \linenumbers

%\tableofcontents
%\listoffigures
%\listoftables
%\newpage

\section{Introduction}
\label{sec:Introduction}

The plastic deformation behavior of face-centred cubic (fcc) single crystalline metals is characterized by the phenomenon of strain hardening, especially in stage II.
During strain hardening, the formation of dislocation networks and the associated hindrance of dislocation mobility play an important role as observed in early experimental studies \cite{livingston_density_1962, basinski_dislocation_1964, pande_dislocation_1971}.
Therefore, it is essential to understand and describe the mutual dislocation interaction and reactions and their effects on the dislocation microstructure.

Taking into account the microstructure evolution, different continuum approaches have been developed to represent the hardening behavior.
The models introduced include physical based formulations using the intrinsic material behaviour and phenomenological motivated formulations using extrinsic properties and observations, e.g. \cite{Kubin2008a,Andreoni_2020,Roters2019}.
One well-known crystal plasticity approach represents the coupling of the plastic slip-controlled dislocation multiplication of a Kocks-Mecking formulation \cite{Kocks2003} with a Taylor- or Franciosi-like yield stress \cite{Taylor1934, Taylor1934a, Franciosi1980} to consider the strain hardening, e.g. \cite{ma_dislocation_2006, Kubin2008, Kubin2008a, alankar_explicit_2012, Reuber.2014, Roters2019}.

The yield stress correlates with the square root of the dislocation density and Taylor uses a scalar prefactor for the yield stress approximation \cite{Taylor1934,Taylor1934a}.
Franciosi et al. found that the prefactor strongly depends on the interaction type between two slip systems and extended the Taylor law by a matrix of slip-system-wise prefactors \cite{Franciosi1980, Franciosi_1982a,Franciosi_1982b}.
Madec et al. \cite{madec_dislocation_2002} confirm the value of the scalar prefactor of the Taylor law, which is estimated by theory, e.g. \cite{Schoeck1972}, as well as by experiment, e.g. \cite{neuhaus_flow_1992}, by performing discrete dislocation dynamic (DDD) simulations.
Based on DDD simulations, the influence of each individual dislocation interaction has been explored by using the Franciosi-like yield stress and its prefactor interaction matrix \cite{madec_role_2003,Devincre2006,Kubin2008,Madec2008}.
Thus, the approach of Franciosi et al. relies on a more profound consideration of the microstructure evolution by considering details of dislocation configurations and interactions comprising reactions.
Other approaches distinguish between different types of dislocation densities, such as edge and screw dislocations \cite{Arsenlis.2002,Reuber.2014,alankar_explicit_2012,leung.2015}, including the curvature of the dislocation lines \cite{Hochrainer2007,schulz_mesoscale_2019}, considering the formation of dislocation dipoles \cite{Reuber.2014, schulz_dislocation-density_2017}, and/or using separate formulations for mobile and immobile dislocations \cite{Ma2004, li_predicting_2014, leung.2015, Sudmanns_2020}.

Recently, continuum dislocation dynamics (CDD) based models have been introduced that describe individual dislocation reactions between the single slip systems.
These are dislocation annihilation due to dislocation climb, cross-slip and collinear reactions \cite{monavari_annihilation_2018,Sudmanns_2020}, dislocation multiplication due to (double) cross-slip and glissile reactions \cite{monavari_annihilation_2018,sudmanns_dislocation_2019} as well as dislocation immobilization due to Lomer reactions and the associated formation of dislocation networks \cite{Sudmanns_2020}. 
In order to ensure the physical basis of the continuum models one depends on  knowledge transfer of the underlying length scales.
A roadmap for scale bridging different simulation approaches with additional information of experimental results is presented in \cite{Giessen2020}.
The arising challenges of connecting different length scales are addressed in \cite{Andreoni_2020}.
Some of the limits of dislocation based continuum theories are discussed in \cite{Schulz2018}.

In order to find meaningful homogenization models for continuum crystal plasticity formulations for dislocation network structures, there is increasing interest in data-driven methods such as machine learning to study the microstructural behavior \cite{Morgan2020, Huber_2020, Steinberger2019}.
Starting from a well-defined data set, these methods can also be applied to predict new data \cite{Salmenjoki2018,Mangal2018}.
Working with DDD simulation data sets of fcc metals under uniaxial tensile loading, discrete distributions of dislocations were homogenized into a dislocation density by \cite{Akhondzadeh2021,Akhondzadeh2020,sills_dislocation_2018,Davoudi2018} to identify aspects of the microstructure evolution, such as the yield stress and dislocation multiplication.
In the paper from Sudmanns et al. \cite{Sudmanns_2020}, analyzes are shown for the reaction densities of individual dislocation reactions between slip systems based on a collision rate \cite{Stricker2015} for a DDD data set of $\langle1\,0\,0\rangle$ oriented aluminum single crystal under uniaxial tensile loading.
The mentioned approaches illustrate the potential of data-driven analysis to identify microstructural characteristics in the transition regime between discrete to continuum formulations.

In this work, we use data science approaches as evaluation method for common domain knowledge such as, e.g., commonly used continuum formulations.
Based on that we extend the formulations to enable a knowledge transfer from discrete to continuum formulations.
We analyze a three-dimensional DDD data set of uniaxial tensile tests of single crystalline aluminum cubes with varying crystal orientation.
Herein, we particularly focus on the evolution of the dislocation network structures and identify a rate formulation for the homogenized dislocation behavior due to different dislocation reactions (glissile, coplanar, cross-slip, collinear and Lomer).
The influence of the slip system activity of the slip systems involved in the dislocation reaction is investigated and considered as key for an averaged description.

The paper is organized as follows: Section \ref{sec:Methods} describes the general workflow, the slip system dependent rate formulation of network evolution, and the data science methods used. 
The systems and data sets considered, including Schmid factors and slip system reactions, are presented in section \ref{sec:System_DataSet}.
The results of the stress-strain response, the dislocation microstructure evolution, as well as determined reaction coefficients and prediction qualities depending on the slip system activities are shown in section \ref{sec:Results}. 
These results are discussed in detail in section \ref{sec:Discussion} with emphasis on the methodology, slip system activity, and reaction coefficients.
The paper concludes with a summary in section \ref{sec:Conclusion}.

%CORRECTION DONE

\section{Methods}
\label{sec:Methods}

We use data analysis to enable a knowledge transfer from discrete to continuum formulations.
The objective is the derivation of interpretable data-driven models, which are fully traceable.
We show that the models are able  to incorporate fundamental physical processes using data science into a physically based modelling approach.

\subsection{Workflow - from discrete to continuum}
\label{sub:workflow}

\begin{figure}
\centering
    \includegraphics[width=0.8\textwidth]{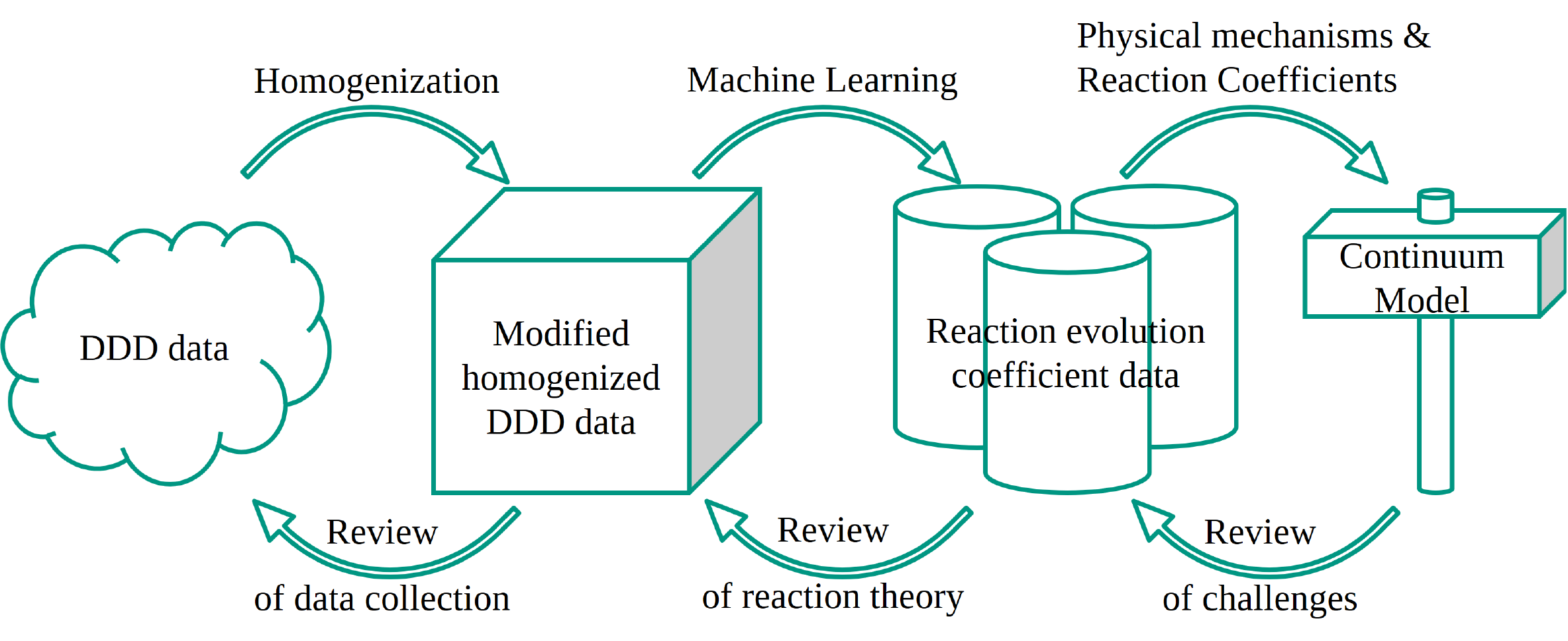}
    \caption{The iterative workflow from a DDD data set to the consideration of identified mechanism within a continuum formulation is shown.
    }
    \label{fig:workflow}
\end{figure}

The general workflow is shown in \autoref{fig:workflow}.
Our initial data was gained by DDD simulations, where dislocation networks evolve during plastic deformation.
The discrete data sets of DDD simulations involve information about the local distribution of dislocations and topological connectivity, based on a graph description of the three-dimensional dislocation network \cite{Weygand_2002,ElAchkar2019} (see \autoref{fig:initial dislocation density} in section \ref{sec:System_DataSet}).
This graph description allows to track deformation history and thus the distribution of plastic slip within the sample.

Homogenization schemes are employed to extract dislocation density representation of the initial discrete DDD data: the discrete dislocation structure is averaged over a defined volume resulting in a dislocation density.
Moreover, we consider further discrete characteristics of DDD data set, i.e. we homogenize the dislocation reactions within the discrete dislocation network into so-called reaction densities, and the swept areas of dislocations, allowing the determine homogenized plastic shear strains at the level of slip systems.
In the context of machine learning, the homogenized characteristics are called features, used here for the predictions.
Additionally, feature engineering is applied, which yields a mathematical operation on single features and on combinations of different features, respectively.
Thus, feature engineering generates further features characterizing the data.
One main function of feature engineering is the transformation of nonlinear into linear model dependencies, e.g. we transform a square rooted feature into a new feature, on which the square root has already been applied.

Based on a set of homogenized features in order to identify and validate relationships and interactions, the next methodological step is applying a machine learning algorithm on the homogenized data by a set of model equations (see arrow "Machine Learning" in \autoref{fig:workflow}).
The model equations are reaction rate equations to investigate the evolution of dislocation networks. 
The reaction evolution equations are derived by domain knowledge which is physically based.
A detailed explanation for the equations is given in \autoref{subsub:uniform_approach}.
The prediction of our non-linear models can be solved by e.g. artificial neural networks or random forests \cite{Forsyth2019, James2013, Unpingco2019}.
However, the non-linear features can be transformed to linear models by feature engineering, to enable using multilinear regression for solving the model equations.
We choose to use linear regression for the prediction in order to reduce the model complexity and to guarantee a high level of interpretability due to our objective of providing a traceable model.
After applying the machine learning model, the regression coefficients, the so-called reaction rate coefficients, are generated, which specify the kinetics of the reaction density evolution.

As indicated on the right-hand side of \autoref{fig:workflow}, the reaction equations and the reaction rate coefficients can be transferred to crystal plasticity continuum models in order to provide a physically based extension to improve on the description of the dislocation network evolution.
The application of the reaction equations in continuum models is beyond the scope of this work, but its application in CDD modelling in the context of a comparison with the existing DDD dataset is straightforward.

Each single procedure step needs to be reviewed carefully (see backward facing arrows in \autoref{fig:workflow} as a feedback loop) and, if necessary, executed iteratively, since there might be unforseen artefacts: 
E.g., the chosen time step for determining the values of the rates of interest may be too small, so that the discrete behavior of the dislocations in DDD does not allow for an adequate continuum description (material science artefact), or the chosen time step is too large, so the data set can shrink too much (data science artefact), which leads to a insufficient amount of data for data science methods.

\subsection{Physically based model for dislocation reaction evolution}
\label{sub:physical_based_model}

In three-dimensional DDD simulations, dislocation networks is highly entangles and complex, comprising junctions and mobile dislocations. 
The dislocation network stabilizes dislocations, which can in principle glide.
While deforming the material, the dislocation network evolves. Thereby, mobile dislocations react with both mobile dislocations or  with immobile dislocations stabilized by junctions.
In some instances, immobile junctions can be re-mobilized \cite{Sudmanns_2020}. 
In this work, we focus on the evolution of dislocation junctions in fcc namely the Lomer, Hirth, glissile, coplanar, collinear and cross-slip using their homogenized characteristics.

\subsubsection*{Plastic shear based model}

Regarding the incorporation of dislocation reaction processes in the continuum model, we start with the consideration of classical phenomenological approaches based on the plastic shear rate of a slip system. 
In some approaches, e.g. Roters et al \cite{Roters2010}, the plastic shear rate $\partial_t \gamma^{\xi}$ is described by a function of the resolved shear stress $\tau^{\xi}$ and the critical resolved shear stress $\tau_{cr}^{\xi}$ on the considered slip system. 
In these formulations, the material state is estimated using a function of the total shear $\gamma$ and the total shear rate $\partial_t \gamma$:
\begin{equation} \label{eq: tau_cr}
    \partial_t \gamma^{\xi} = f(\tau^{\xi},\tau_{cr}^{\xi}),\quad \tau_{cr}^{\xi} = f(\gamma,\partial_t \gamma)
\end{equation}
A shortcoming is that the material state is solely dependent on the critical resolved shear stress  $\tau_{cr}^{\xi}$.
A different approach was already proposed by Taylor \cite{Taylor1934,Taylor1934a} to account for work hardening.
Thereby, the critical resolved shear stress and dislocation density $\rho$ are related by $\tau_{cr}^{\xi} \propto \sqrt{\rho}$.
Kocks and Mecking formulated a dislocation density model for work hardening, which is based on the theory of dislocation accumulation \cite{Kocks2003}:
\begin{equation} \label{eq: KocksMecking rate}
    \frac{d\rho}{d\gamma} = \frac{1}{b}\frac{dL}{dA}.
\end{equation}
Hereby, $b$ is the length of the Burgers vector and $ dL/dA$ is a measure for the dislocation length $dL$ stored after sweeping an area~$dA$.
This quotient can be approximated with $dL/dA \propto \sqrt{\rho}$, which derives from the reciprocal of the average dislocation spacing. 
Consequently, the multiplication term provides the dislocation density as a function of total shear, which can be substituted into the Taylor hardening term.
The phenomenological models are easily accessible, since external parameters are more easy to measure than internal ones \cite{Roters2010}.
However, the microstructural behaviour and size-dependent effects are not considered, what limits this kind of models.

\subsubsection*{Dislocation density based model}

Physically based models usually rely on internal properties such as dislocation density $\rho^{\xi}$ on slip system~$\xi$.
The Orowan equation is a physical based relation between the kinetic of mobile dislocation density $\rho_M^{\xi}$ and the resulting plastic slip rate~$\partial_t \gamma^{\xi}$ based on the average dislocation velocity~$\upsilon^{\xi}$ \cite{Orowan1934a,Orowan1934b,Orowan1934c}:
\begin{equation} \label{eq: orowan}
    \partial_t \gamma^{\xi} = \upsilon^{\xi} b \rho_M^\xi.
\end{equation}
Considering dislocation networks, the dislocation density $\rho^{\xi}$ can be divided into a mobile and a network dislocation density.
The mobile dislocation density $\rho_M^{\xi}$ is crucial for plastic deformation.
The network dislocation density $\rho_{net}^{\xi}$ is important for hardening.
Here it is subdivided into a Lomer density~$\rho_{lom}^{\xi}$ and a stabilized dislocation density~$\rho_S^{\xi}$.
An overview is given in Zoller et al.~\cite{Zoller2021}. However, in DDD, it is not trivial to get this exact information about the different types of densities.
Furthermore, in DDD the velocity has a jagged behaviour, since dislocations can glide during one averaging time step and stop at the next time step due to dislocations interactions.
For continuum approaches, this behaviour can be averaged out by using larger averaging time steps or the swept area $A^{\xi}$, which has been found to be a more robust parameter and allows to calculated the plastic shear rate on slip system ${\xi}$ as:
\begin{equation} \label{eq: shear_rate_A}
    \partial_t \gamma^{\xi} = \frac{b}{V} \partial_t A^{\xi}.
\end{equation}
By combining \autoref{eq: orowan} and \autoref{eq: shear_rate_A}, a simple relation for the dislocation velocity is obtained.
The total plastic strain tensor $\boldsymbol{\epsilon}_{pl}$ of the considered volume is given by
\begin{equation} \label{eq: plastic strain tensor}
    \boldsymbol{\epsilon}_{pl} = \sum_{\xi=1}^{12} \frac{A^{\xi}}{2V}\left( \boldsymbol{b}^{\xi} \otimes \boldsymbol{n}^{\xi} +  \boldsymbol{n}^{\xi} \otimes \boldsymbol{b}^{\xi}\right),
\end{equation}
where $\boldsymbol{n}^{\xi}$ is the slip plane normal and $\boldsymbol{b}^{\xi}$ is the Burgers vector of slip system~${\xi}$.

Starting from a slip system specific formulation analogous to \autoref{eq: KocksMecking rate} including the approximation by the square root of the dislocation density, the combination with the Orowan equation (\autoref{eq: orowan}) provides a rate equation of the dislocation densities on the individual slip systems with the following form:
\begin{equation} \label{eq: proportional reaction rate}
    \partial_t \rho^{\xi} \propto \frac{1}{b}\sqrt{\rho} \,\partial_t \gamma^{\xi}
    \quad\rightarrow\quad
    \partial_t \rho^{\xi} \propto \upsilon^{\xi} \rho_M^\xi \sqrt{\rho},
\end{equation}
which is also motivated by the product of the mobile dislocation density with a dislocation collision rate~$\phi^{\xi \rightarrow \zeta} = f(\xi, \zeta)$ between the slip systems $\xi$ and $\zeta$:
\begin{equation} \label{eq: collision rate}
    \phi^{\xi \rightarrow \zeta} = \frac{\upsilon^{\xi}}{L^{\zeta}}.
\end{equation}
Thereby, the average dislocation spacing $L^{\zeta} \propto 1/\sqrt{\rho^{\zeta}}$ is used.
The collision rate approach was introduced in \cite{Ma2004} and was used in \cite{Kubin2008a,Stricker2015,Roters2019}.
In this work, we use this relation and extend it for modelling reaction densities, which is explained in detail in the following.

\subsubsection{Uniform approach for network evolution}
\label{subsub:uniform_approach}

A reaction rate equation was introduced in~\cite{Stricker2015, Roters2019} and was applied to the dislocation network formulation in \cite{Sudmanns_2020}.
These formulations are based on the assumption that the reaction density rate~$\partial_t \rho_{react}$ is due to two interacting slip systems with a collision rate~(\autoref{eq: collision rate}), which leads to the following relation:
\begin{equation}
    \partial_t \rho_{react} \propto \rho_M^{\xi}\phi^{\xi \rightarrow \zeta} + \rho_M^{\zeta}\phi^{\zeta \rightarrow \xi}
    \quad \text{with} \quad \xi \neq \zeta.
\end{equation}
This leads to the following reaction rate equation:
\begin{equation} \label{reaction_denisity_derivative_rate}
    \partial_t\rho_{react} = C_{react} \left( \frac{1}{b} \left| \partial_t \gamma^\xi \right| \sqrt{\rho_M^\zeta + \rho_S^\zeta} + \frac{1}{b} \left| \partial_t \gamma^\zeta \right| \sqrt{\rho_M^\xi + \rho_S^\xi} \right)
\end{equation}
The subscript $()_{react}$ indicates all types of reactions (cp. \autoref{fig:interaction_matrix}), except for the self-interaction, but it includes in this work also the cross-slip mechanism, which is topologically handled as a dislocation junction with zero Burgers vector. 
It should be noted that the cross-slip mechanism is physically not based on the collision of two dislocations.
However, in order to apply a general approach and be aware of possible misinterpretations of data-driven results, cross-slip is included in the analysis of the dislocation network.
From the topological point of view, the collinear and cross-slip junctions have the same properties: connecting two slip systems which share the Burgers vector.
$C_{react}$ is a constant reaction dependent coefficient.
The dislocation velocity is replaced by the Orowan equation (\autoref{eq: orowan}) with the plastic shear rate.
In addition to the mobile dislocation density, the stabilized dislocation density contributes to all dislocation reactions and therefore to the average dislocation spacing.
This leads to the assumption that the mean distance of the interacting slip system $\zeta$ is $1 / L^{\zeta} = \sqrt{ \smash[b]{\rho_M^{\zeta}} + \smash[b]{\rho_S^{\zeta}} }$. 
With respect to the total dislocation density, which consists of mobile, stabilized and Lomer dislocation density\footnote{We neglect the Hirth dislocation density due to the very rare event of this reaction type in the considered data set, which is in accordance to \cite{sills_dislocation_2018}}, we rewrite the average dislocation spacing to $\sqrt{ \rho_{M}^{\zeta} + \rho_{S}^{\zeta}} = \sqrt{\rho_{tot}^{\zeta} - \rho_{lom}^{\zeta}}$.
Hereby, it is remarked, that for a mobile dislocation density on a slip system, the Lomer density on another slip system can not be partner for any dislocation reaction.

Different levels of homogenization are applied based on the reaction rate \autoref{reaction_denisity_derivative_rate}.
The homogenization for each model relates to the considered number of slip systems.
Thus, we introduce a \textit{individual}, a \textit{summed} and a \textit{grouped} model in the following, which differ in the number of constant coefficients describing the total dislocation network evolution.

The first \textit{individual} approach aims to predict the reaction density for each slip system interaction pair individually, i.e. a reaction coefficient~$C_{react}^{\xi,\zeta}$ is calculated for each slip system interaction between~$\xi$ and~$\zeta$.
The scale of homogenization is very fine, since each interaction is treated individually, which leads to a modification of \autoref{reaction_denisity_derivative_rate} to following  equation:
\begin{equation} \label{eq: reaction denisity rate individual}
    \partial_t\rho_{react}^{\alpha} = \sum_{(\xi,\zeta)} C_{react}^{\xi,\zeta} \left( \frac{1}{b} \left| \partial_t \gamma^\xi \right| \sqrt{\rho_{tot}^\zeta - \rho_{lom}^\zeta} + \frac{1}{b} \left| \partial_t \gamma^\zeta \right| \sqrt{\rho_{tot}^\xi - \rho_{lom}^\xi} \right)
\end{equation}
The interaction of the slip systems lead to a reaction density on slip system~$\alpha$, which is not necessarily equal to~$\xi$ or~$\zeta$.
The reaction slip system depends on the type of reaction, which is described in \autoref{sub:reaction_types_DDD}.
This leads to a symmetric 12 $\times$ 12 reaction coefficient matrix with 66 different coefficients due to the symmetry and neglecting the self-interaction for the \textit{individual} approach.

For a coarse homogenization approach, we focus on the behaviour of a set of interacting slip systems.
Thereby, each set relates to the type of reaction, i.e., it consists of the interacting slip systems $\xi$ and $\zeta$, which store density of a specific type of reaction.
Additionally, we do not distinguish, on which slip system the reaction density is deposited, i.e. we sum over all slip systems $\alpha$ for the reaction density.
Based on these assumptions, we modify \autoref{reaction_denisity_derivative_rate} to the \textit{summed} equation:
\begin{equation} \label{eq: reaction denisity rate summed}
    \sum_{\alpha=1}^{12} \partial_t\rho_{react}^{\alpha} = C_{react} \sum_{(\xi,\zeta)} \left( \frac{1}{b} \left| \partial_t \gamma^\xi \right| \sqrt{\rho_{tot}^\zeta - \rho_{lom}^\zeta} + \frac{1}{b} \left| \partial_t \gamma^\zeta \right| \sqrt{\rho_{tot}^\xi - \rho_{lom}^\xi} \right)
\end{equation}
Hereby, each type of reaction is limited to one slip system independent coefficient $C_{react}^{\xi,\zeta} \rightarrow C_{react}$.
Therefore, this {summed} approach is a coarse homogenization, since there is only one reaction coefficient for each reaction type.

The third approach is on a scale between the fine homogenization of \autoref{eq: reaction denisity rate individual} and the coarse homogenization of \autoref{eq: reaction denisity rate summed} and matters for glissile and Lomer reactions.
For the other types of reactions there is no difference to the \textit{individual} approach in \autoref{eq: reaction denisity rate individual}.
Thereby, groups of slip systems are considered for the reaction rate equation.
The groups are derived based on two aspects.
Firstly, for the glissile reaction there is more than one possible 
reaction, which leads to a reaction density on slip system~$\alpha$.
Therefore, we group all four interacting slip systems, which deposit reaction density on $\alpha$.
Secondly for the Lomer reaction, the reaction product is counted on the two reaction slip systems by 50$\%$. 
Since for our considered data set only the total resulting reaction density on the considered slip system was used, a precise allocation into the contributions of the individual reaction pairings is not possible in this work.
However, there is the peculiarity of Lomer reactions, that the summed reaction density of three slip systems $\alpha$ can be traced back to various reactions between two slip systems $\xi$ and $\zeta$ each, both belonging to the group of these three slip systems (see \autoref{sub:reaction_types_DDD}, e.g. slip system group \{A6,B4,C1\}). 
The two modifications for glissile and Lomer reactions leads to the \textit{grouped} equation
\begin{equation} \label{eq: reaction denisity rate group}
    \sum_{\alpha \in A} \partial_t\rho_{react}^{\alpha} = C_{react}^{j} \sum_{(\xi,\zeta) \in B}  \left( \frac{1}{b} \left| \partial_t \gamma^\xi \right| \sqrt{\rho_{tot}^\zeta - \rho_{lom}^\zeta} + \frac{1}{b} \left| \partial_t \gamma^\zeta \right| \sqrt{\rho_{tot}^\xi - \rho_{lom}^\xi} \right).
\end{equation}
Hereby, each type of reaction has a number of groups $j$, which consist of a group of reaction slip systems $\alpha \in A$ and of a group of interacting slip system pairs $(\xi,\zeta) \in B$, where $A$ is a set of reaction slip systems for the interacting sets of slip systems $B$.
The glissile reaction is divided into $j$ = 12 groups with only one system $\alpha$ per group $A$.
The Lomer reaction is divided into 4 groups, where each group $j$ consists of three reaction slip systems $\alpha$ in $A$ and three interacting slip systems pairs $(\xi,\zeta)$ in $B$.
Therefore, one reaction coefficient $C_{react}^{j}$ results for each group $j$, which leads to a group of reaction coefficients for each type of reaction.
A summary of the three different approaches for \autoref{eq: reaction denisity rate individual}, \autoref{eq: reaction denisity rate summed} and \autoref{eq: reaction denisity rate group} is given in \autoref{tab:group_coefficient}. 

{
\renewcommand{\arraystretch}{1.1}
\begin{table}
    \centering
    \caption{List of groups and amount of coefficients for the \textit{individual}, \textit{summed} and \textit{grouped} model equations. In the right column the number of coefficients for the activity dependent post-analysis approach is given. The (*) indicates, where the Hirth reaction is not calculated.}
    \begin{tabular}{m{0.1\textwidth}>{\centering}m{0.19\textwidth}>{\centering}m{0.19\textwidth}>{\centering}m{0.19\textwidth}>{\centering\arraybackslash}m{0.19\textwidth}}
        \toprule[1pt]\midrule[0.3pt]
        & \multicolumn{4}{c}{Number of constant coefficients by approach:} \\
        &    \textit{individual}  &  \textit{summed} &   \textit{grouped} &   activity dependent \\
        \hline
        Glissile        &   24  &   1   &   12  & 3\\
        Collinear       &   6  &   1   &   6    & 3\\
        Coplanar        &   12  &   1   &   12  & 3\\
        Lomer           &   12  &   1   &   4   & 3\\
        Hirth           &   12*  &   1 &   12*    & 3*\\
        \hline
        Total           &   66  &   5   &   44  & 15\\
        \midrule[0.3pt]\toprule[1pt]
    \end{tabular}
    \label{tab:group_coefficient}
\end{table}
}

\subsubsection{Schmid factor dependent analysis}
\label{subsub:schmid_factor_dependent}

Based on the reaction rate equations in \autoref{subsub:uniform_approach}, we investigate the impact of the slip system activity on the resulting reaction coefficients.
This approach is an activity-dependent homogenization.
Thereby, the information about the activity of a slip system is estimated by the Schmid factor of each slip system.

In this work, there is a clear separation between inactive and active for the highly symmetric orientations $\langle 100 \rangle$, $\langle 110 \rangle$ and $\langle 111 \rangle$, whereby all active slip systems have the same Schmid factor.
In contrast, the activity of the slip systems differs for the $\langle 123 \rangle$ orientation (cp. \autoref{tab:Schmidfactor}). 
With reference to the resulting plastic shear (see \autoref{fig:shear_strain} in the Appendix) and under the premise that the chosen limit is also valid for the active slip systems in the $\langle111\rangle$ orientation, we use a minimum Schmid factor $S_{min}$ of 0.25 for the separation between active and inactive, i.e. there are four active and eight inactive slip systems for the $\langle 123 \rangle$ orientation. 
For our approach, the interacting slip systems $\xi$ and $\zeta$ are either both inactive or both active or one is active and the other one is inactive.
This conditional clustering by activity subdivides the reaction coefficient into three cases: 
\begin{equation} \label{eq: coefficient case group}
    C_{react}=\begin{cases}
        \frac{1}{n} \sum_{(\xi,\zeta)}^n C_{react}^{\xi,\zeta}, & \text{if $(\xi \wedge \zeta) \in S \geq S_{min}$}.\\
        \frac{1}{n} \sum_{(\xi,\zeta)}^n C_{react}^{\xi,\zeta}, & \text{if $(\xi \oplus \zeta) \in S \geq S_{min}$}.\\
        \frac{1}{n} \sum_{(\xi,\zeta)}^n C_{react}^{\xi,\zeta}, & \text{if $(\xi \wedge \zeta) \notin S \geq S_{min}$}.
    \end{cases}
\end{equation}
The reaction coefficient $C_{react}$ for each type of activity is calculated by the mean of involved $n$ reaction coefficients $C_{react}^{\xi,\zeta}$.
This leads to three activity dependent coefficients for each reaction type (see \autoref{tab:group_coefficient}).

\subsubsection{Regression modelling}
\label{subsub:regression_model}

In this work, we solve the reaction density rate equations of \autoref{subsub:uniform_approach} by multi linear regression models.
In the multiple linear regression (MLR) model, there are linear models in all features.
Therefore, the model has a linear dependency of the target with different coefficients, features and an error term, which has the form $\boldsymbol{y} = \boldsymbol{X} \boldsymbol{\beta} + \boldsymbol{\epsilon}$, where $\boldsymbol{y}$ are the target values, $\boldsymbol{X}$ is the matrix of features and $\boldsymbol{\epsilon}$ is a constant error, e.g. the bias of the prediction.
The coefficients $\boldsymbol{\beta}$ are estimated by a least-square estimator.
The quality of the regression model is the coefficient of determination $R^2$, which has a range $\in (-\infty, 1]$, since the error term $\boldsymbol{\epsilon}$ is set to zero 
\footnote{
Remark for coefficient of determination $R^2$: The coefficient of determination is an indicator of how well the model predicts the target data.
If MLR is applied with a non-null error term $\boldsymbol{\epsilon}$ then $R^2 \in [0, 1]$.
If the error term $\boldsymbol{\epsilon}$ is fixed to a zero vector, then $R^2 \in (-\infty, 1]$, since the prediction can be worse then the mean target value.
}. 
We choose a zero error term, since we evaluate our physical based formulations of \autoref{subsub:uniform_approach}, which do not contain any additional term.
In our model, the reaction density rates are used as the target values and the variables of the right-hand side of the reaction rate equations are treated as the features.

It is of great importance in the field of physics-based analysis of dislocations, that the way of modelling is traceable.
This can not be provided by a more complex approach in data science like neuronal networks, whose algorithm leads to black box models, which is a non-parametric approach.
Thus, physical theories can not be proven adequately.
Therefore, we use linear regression models, which are parametric approaches.
Thus, it leads to a traceable and introspective models, so-called knowledge-based models or \textit{white box} models, respectively.

%CORRECTION DONE

\section{System and data set}
\label{sec:System_DataSet}

Based on the DDD code described in \cite{Weygand_2001,Weygand_2002,Weygand2005, ElAchkar2019}, simulations of tensile tests on fcc single crystalline aluminum have been performed.
The simulations are carried out six times each for four crystal orientations which leads to 24 DDD simulation data sets. 
The crystal orientations considered are $\langle 100 \rangle$, $\langle 110 \rangle$, $\langle 111 \rangle$, and $\langle 123 \rangle$.
The orientations $\langle 100 \rangle$, $\langle 110 \rangle$, and $\langle 111 \rangle$ are high symmetry multi-slip orientations, while orientation $\langle 123 \rangle$ is a single slip (non-high symmetry) orientation.

\begin{figure}
    \centering
    \begin{subfigure}{0.48\textwidth}
    \includegraphics[width=0.44\linewidth]{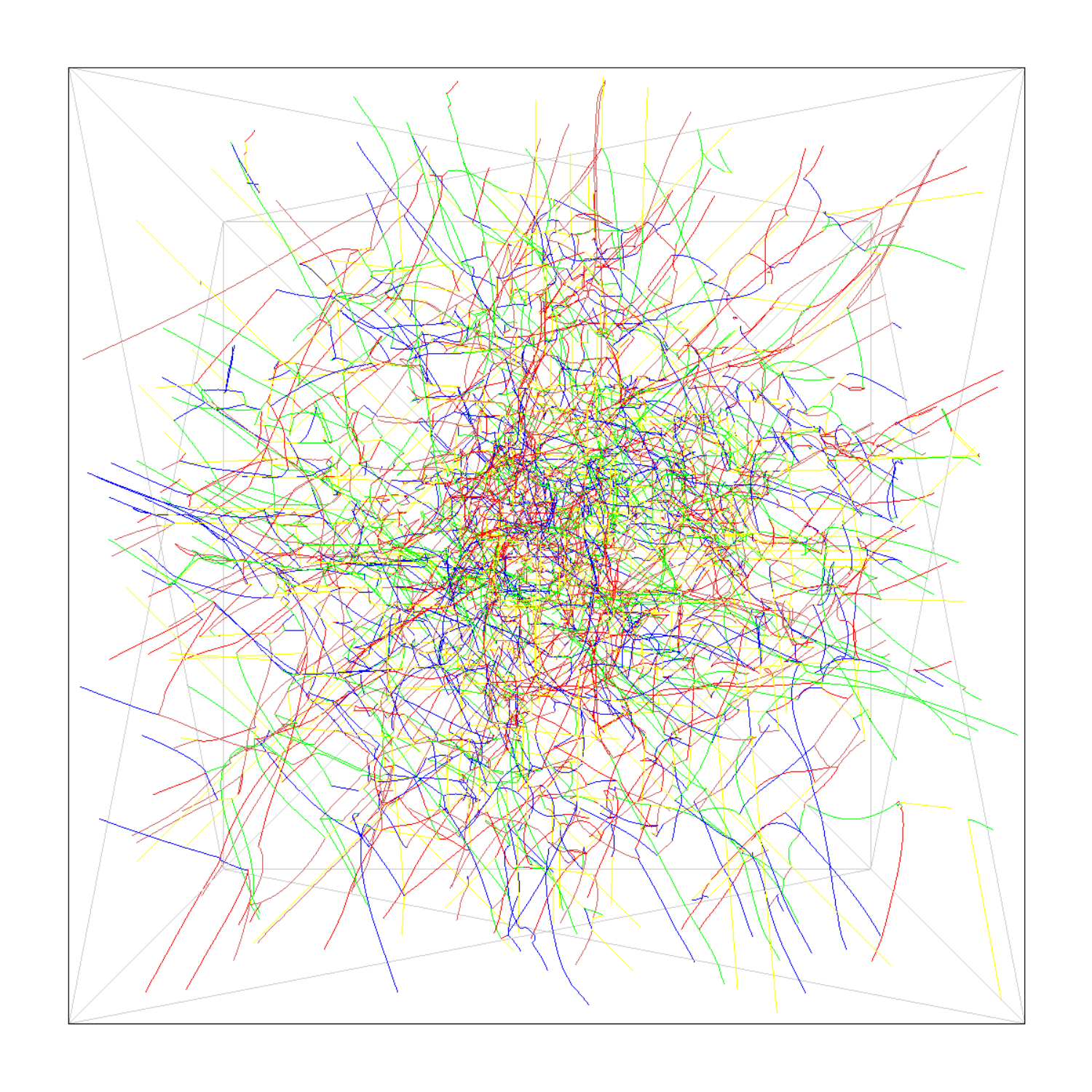}
    \includegraphics[width=0.52\linewidth]{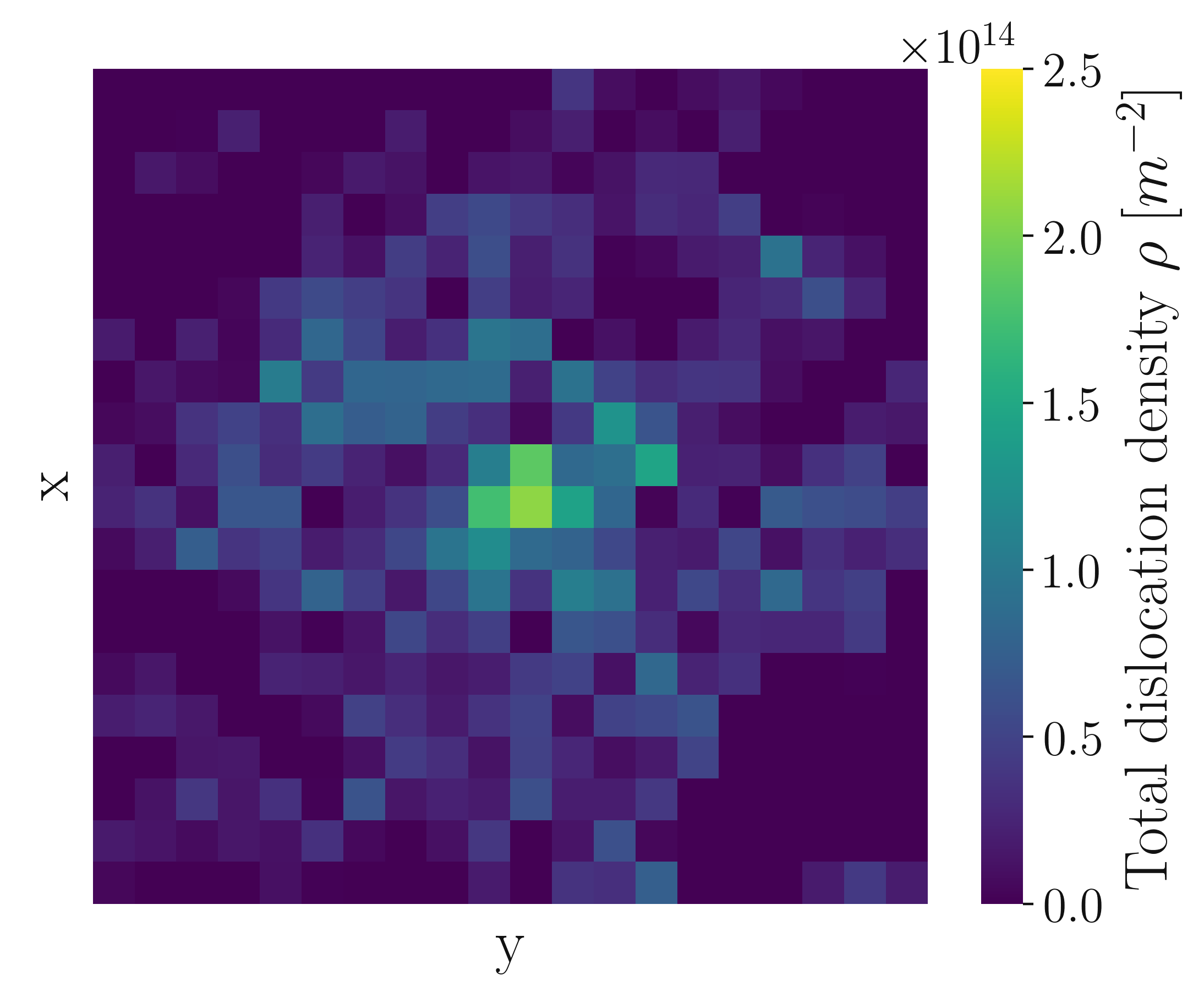}
    \caption{Initial relaxed configuration at $\epsilon_{tot} = 0.0\%$}
    \end{subfigure}
    \begin{subfigure}{0.48\textwidth}
    \includegraphics[width=0.44\linewidth]{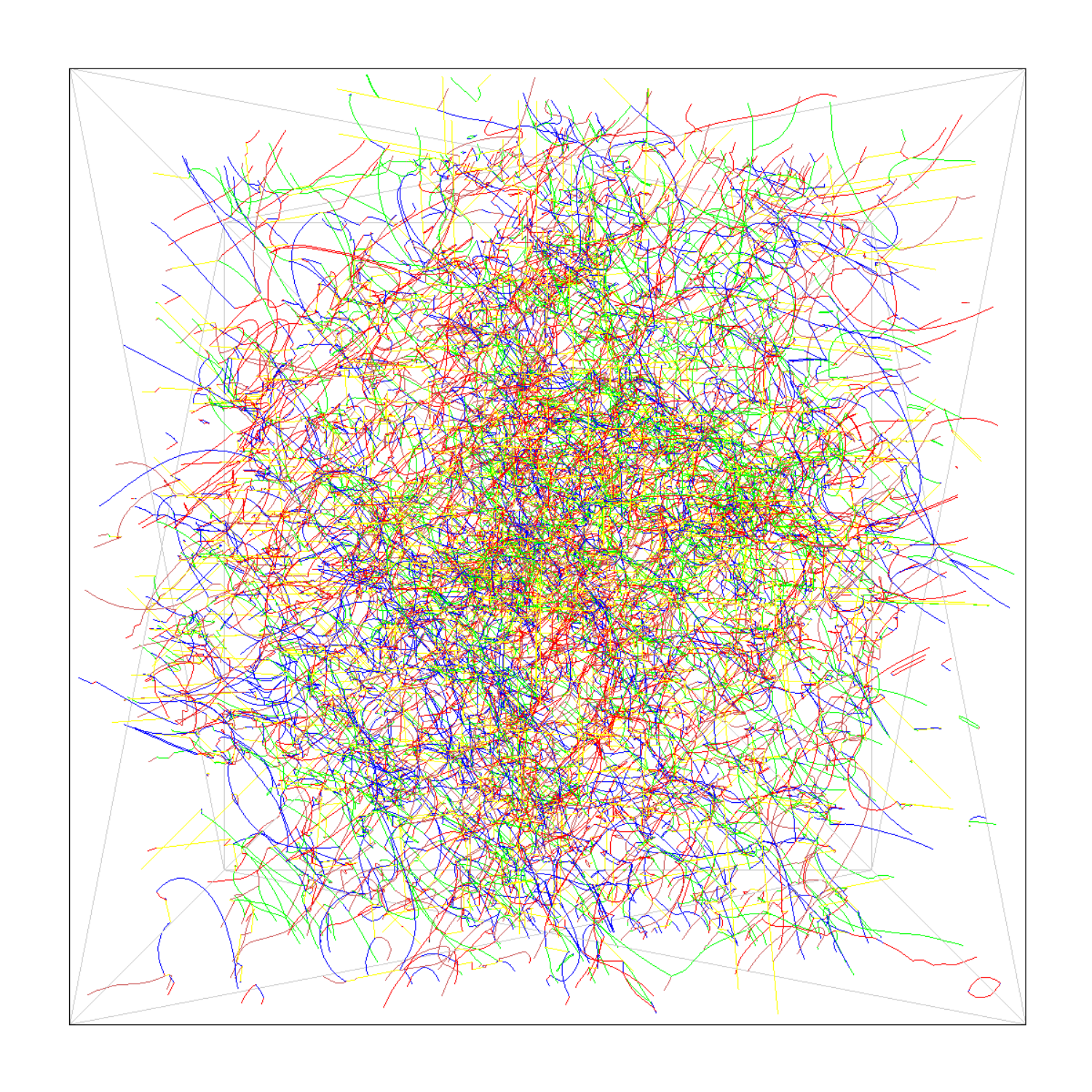}
    \includegraphics[width=0.52\linewidth]{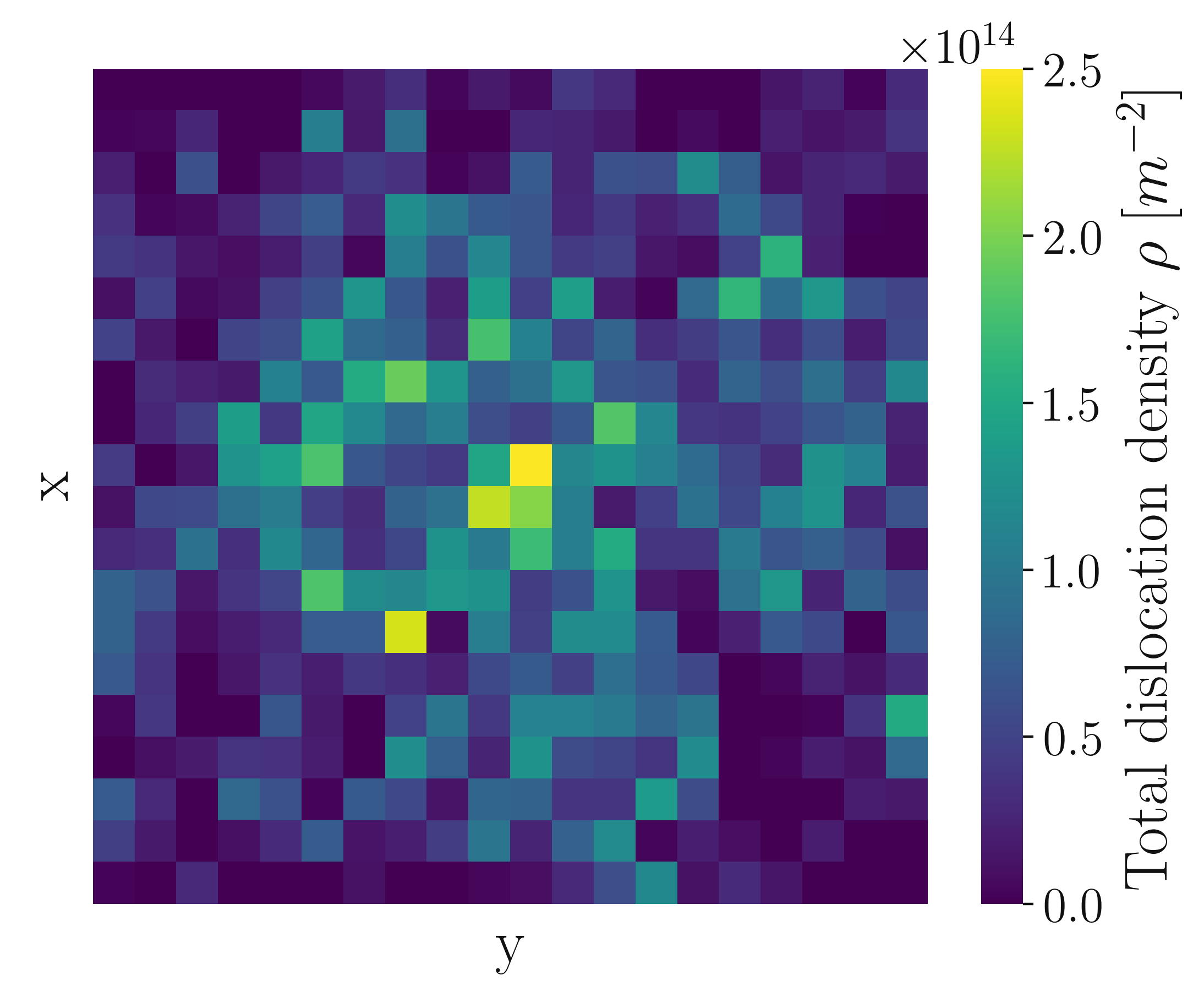}
    \caption{Dislocation configuration at $\epsilon_{tot} = 0.5\%$}
    \end{subfigure}
    \caption{Spatial dislocation distribution in the original DDD data set (respectively left) and the homogenized dislocation densities in the central x-y cross section at $z=2.5 \,\mathrm{\text{\textmu} m}$ (respectively right) for one data set in $\langle 100 \rangle$ orientation of (a) the initial relaxed configuration and of (b) the dislocation configuration at $\epsilon_{tot} = 0.5\%$.}
    \label{fig:initial dislocation density}
\end{figure}

For the preparation of the data, we homogenize the discrete characteristics within the system by deriving continuum variables, like e.g.  the dislocation density (see \autoref{sec:Methods}).  
Discrete structures and their homogenized dislocation density distribution is shown in~\autoref{fig:initial dislocation density}.
In the remained of the paper the averages over the total system volume are used for the analysis of the reaction density.

\subsection{Geometry and initial set up of DDD simulations}
\label{sub:geometry_setup}

The following properties are used for aluminum: lattice constant $a \approx 0.404\,\mathrm{nm}$; shear modulus $\mu = 27\,\mathrm{GPa}$; Poisson ratio $\nu = 0.347$. 
A drag coefficient of $D = 10^{-4}\,\mathrm{Pa \cdot s}$ is chosen.
The simulation framework allows for cross-slip.
The cubic specimen's volume is $(5 \,\mathrm{\text{\textmu} m})^3$.
Regarding the boundary conditions, the displacement of the bottom surface of the cube in tensile direction (y-direction) is fixed.
Additionally the rotational degrees of freedom of the sample are fixed.
At the top surface the displacement in $y$ is prescribed corresponding to an applied total strain rate of $\dot{\epsilon}_{tot} = 5000 \, \mathrm{s^{-1}}$.
The remaining d.o.f. on the top and bottom surfaces ($x$ and $z$ direction) are traction free.
The total strain $\epsilon_{tot}$ reaches up to $0.3 \%$ for the $\langle 110 \rangle$ and $\langle 111 \rangle$ orientations and up to $0.5 \%$ for the $\langle 100 \rangle$ and $\langle 123 \rangle$ orientations.
The DDD framework \cite{Weygand_2002} uses a sub-incremental scheme.
The dislocation structure evolves over many sub-time steps and only every 100 to 500 sub-time steps leading to the time step used from hereon the boundary conditions are updated.

The Schmid-Boas notation for the fcc slip system is used.
The letters A, B, C and D stand for the slip plane normal.
The numbers 1 to 6 mark the possible Burgers vectors in fcc \cite{Schmid1935} as shown in \autoref{tab: Schmid-Boas} in the Appendix.
Additionally, we use the DDD numbering of the slip systems from 1 to 12.
In \autoref{tab:Schmidfactor}, the DDD numbering and the Schmid-Boas numbering are contrasted for a better understanding.

The activity of each of the twelve slip systems depends on the crystal orientation and is described by the Schmid factor $S^{\xi}$.
\autoref{tab:Schmidfactor} shows the Schmid factors for all investigated orientations. 
For the high symmetry orientations, in $\langle 100 \rangle$ there are eight, in $\langle 111 \rangle$ there are six, and in $\langle 110 \rangle$ there are four active slip systems.
For high symmetry orientations, the Schmid factors are equal on active slip systems and zero for inactive slip systems.
Nine slip systems are non-null in the non-high symmetry orientation $\langle 123 \rangle$, but with unequal Schmid factors.
The DDD framework assumes small strain thus no crystal rotation and change in Schmid factor occur.
Consequently, the constant Schmid factors can be calculated a priori. 

For an ideal external tensile stress in y direction ($\boldsymbol{\sigma} = \sigma_{yy} \, \mathbf{e_y} \otimes \mathbf{e_y}$), the Schmid factors are calculated by $ \left|\tau^\xi\right| = \left|\mathbf{M^\xi} \cdot \boldsymbol{\sigma}\right| = S^\xi\,\left|\sigma_{yy}\right|$.
Here, $\mathbf{M^\xi}$ denotes the Schmid tensor defined by the dyadic product of the slip direction $\mathbf{d^\xi}=\frac{1}{b}\,\mathbf{b^\xi}$ and the slip plane normal: $\mathbf{M^\xi} = \mathbf{d^\xi} \otimes \mathbf{n^\xi}$.

The initial dislocation microstructures used for the tensile tests were obtained by a relaxation procedure established in \cite{Motz2009,Stricker2015}, aiming at structures free of artificial pinning points e.g. through Frank-Read sources.
The relaxation procedure was started with randomly distributed circular loops of various size which may also be partly outside the volume (so called virtual dislocations).
The structures relaxed using traction free boundary conditions and a highly interconnected dislocation networks are obtained.
The relaxation was stopped after reaching a constant dislocation density.
The initial relaxed dislocation density $\rho_0$ is in the range of  $1.0 - 1.5 \times 10^{13}\,\mathrm{m^{-2}}$ for the 24 data sets. 

{
\renewcommand{\arraystretch}{1.1}
\setlength\tabcolsep{4pt}
\begin{table}
    \centering
    \caption{Schmid factors of $\langle 100 \rangle$, $\langle 110 \rangle$, $\langle 111 \rangle$ and $\langle 123 \rangle$ orientation for each slip system. The Schmid-Boas notation of the slip systems is indicated with $\mathbf{SB}$ and the DDD numbering is indicated with $\mathbf{i_{DDD}}$.}
    \begin{tabular}{>{\centering}p{0.075\textwidth}*{12}r}
        \toprule[1pt]\midrule[0.3pt]
        \textbf{SB} & A6 & A2 & A3 & B4 & B5 & B2 & C1 & C5 & C3 & D4 & D1 & D6 \\
        $\mathbf{i_{DDD}}$ & 1 & 2 & 3 & 4 & 5 & 6 & 7 & 8 & 9 & 10 & 11 & 12 \\
        \hline
        $\mathbf{\langle 100 \rangle}$ & $0.41$ & $0.41$ &  $0.00$ & 	    $0.00$ & $0.41$ & $0.41$ &         $0.41$ & $0.41$ &  $0.00$        & $0.00$ &  $0.41$ &  $0.41$ \\
        $\mathbf{\langle 110 \rangle}$ &  $0.00$ & $0.41$ &  $0.41$ & 	    $0.00$ & $0.00$ & $0.00$ &         $0.00$ &  $0.00$ &  $0.00$         & $0.41$ &  $0.41$ &  $0.00$ \\
        $\mathbf{\langle 111 \rangle}$ & $0.27$ &  $0.00$ & $0.27$ &         $0.00$ & $0.00$ & $0.00$ & 	    $0.27$ &  $0.00$ & $0.27$        & $0.00$ & $0.27$ & $0.27$ \\
        $\mathbf{\langle 123 \rangle}$ & $0.35$ & $0.12$  & $0.47$ & $0.35$  & $0.17$  & $0.17$ & $0.00$  & $0.00$  & $0.00$  & $0.12$ & $0.29$ & $0.17$ \\
        \midrule[0.3pt]\toprule[1pt]
    \end{tabular}
    \label{tab:Schmidfactor}
\end{table}
}

\subsection{Reaction types in dislocation networks}
\label{sub:reaction_types_DDD}

The data structure of the DDD model can be described as a graph: there are nodes which are connected to segments, representing a straight dislocation section.
There are two type of nodes, the first kind is topological necessary and connects to more than two segments and the second type connects to two segments only and served mainly to discretize (curved) dislocation lines.
Moving dislocation segments may collide on form junctions, which requires reconnecting the dislocation lines at nodes of the first kind.
A dislocation junction is bounded by nodes of the first kind.
A junction consists of a superposition of segments: each loop involved in the junction contributes one segment.
This data structure allows to determine the plastic slip due to each dislocation of a configuration with respect to a reference configuration, e.g. the initial configuration.
We focus on binary reactions, thus, ternary and more complex reactions are neglected in our work \cite{Stricker2015,Stricker_2018}. 
In our DDD framework, the dislocation junction with finite resp. zero Burgers vector are named physical resp. virtual junctions \cite{Weygand_2002}.
\begin{enumerate}
    \item Physical junctions have a finite Burgers vector: in fcc these are the Lomer and Hirth lock.
    Both are sessile and their respective endnodes may only slide along the intersection line of the two glide planes.
    The reaction length is attributed to the two interacting slip systems.
    The reaction density for Lomer and Hirth reactions is assigned equally by $50\%$ each to the two interacting slip system. 
    
    \item Virtual junctions have a zero net Burgers vector.
    They serve to track the topology and the virtual segment length is distance between the respective endnodes.
    Glissile, collinear and coplanar junctions are treated as virtual junctions.
    For a glissile junction a segment of a new glide loop is added to the junction during formation.
    The collinear junction captures the annihilation of two dislocations.
    The coplanar junctions involves two dislocations on the same slip plane but different Burgers vector leading to a new glide dislocations within the plane.
    The concept of virtual dislocation and junctions is detailed in \cite{Weygand_2002,Stricker_2018}.
\end{enumerate}
In addition, the cross-slip mechanism is included in the investigation and measured similarly to the virtual junction approach.
The distance between the start and end point of the cross-slip sector is determined.

\begin{figure}
\centering
    \includegraphics[width=0.35\textwidth]{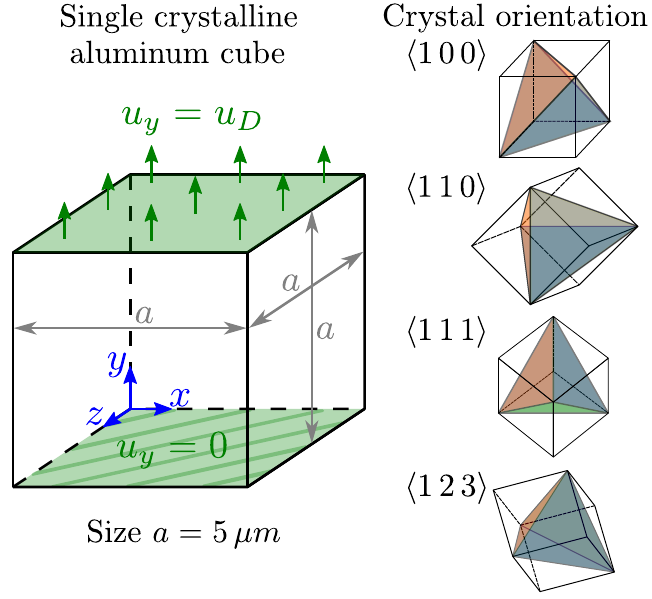}
    \hspace{0.5cm}
    \includegraphics[width=0.55\textwidth]{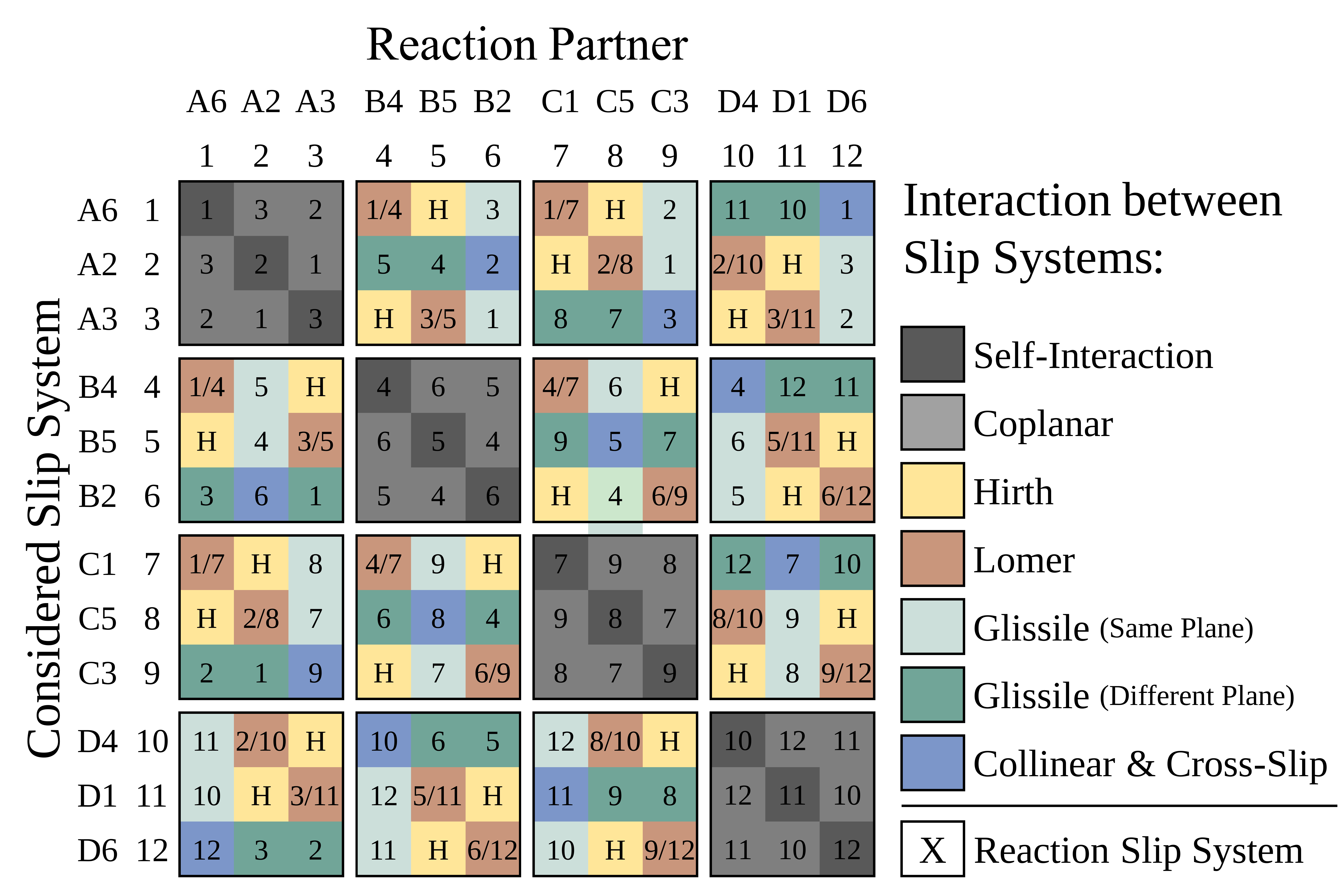}
    \caption{On the left the simulation volume is shown. In the centre the four investigated crystal orientations $\langle 100 \rangle$, $\langle 110 \rangle$, $\langle 111 \rangle$ and $\langle 123 \rangle$ are shown. On the right the interaction matrix for the twelve slip systems in Schmid Boas notation (and its number) for a fcc crystal is shown. 
    The interaction matrix is symmetrical. 
    The different coloration of the glissile reactions has a purely visual aspect in order to recognize the slip systems involved more easily.
    Within the cells the slip system of the reaction product is noted.
    }
    \label{fig:interaction_matrix}
\end{figure}

\autoref{fig:interaction_matrix} shows the interaction matrix for interacting slip systems, whereby the slip system of the reaction product is written in the cells of the matrix.
The color coding relates to the reaction mechanism.
The glissile reaction is split up into reactions, whose reaction product lies on the same plane as the considered slip system and whose product lies on the partner's plane, respectively.
Due to the rare occurrence of the Hirth reaction (energetically not very favourable), we do not investigate Hirth reaction in detail.

%CORRECTION DONE

\section{Results}
\label{sec:Results}

The results are presented as follows.
First, a general overview of the DDD simulation results is given in \autoref{sub:DDD_simulation_results} followed by the results of the dislocation reaction evolution model in \autoref{sub:slip_system_interaction}.
Based on the model results we apply an activity dependent approach, whose findings are presented in \autoref{sub:Activity dependent model results}.
An in-depth analysis of the activity dependency is shown by individual specific Lomer and glissile reactions in \autoref{sub:detailed_model_investigation}.

\subsection{DDD simulation results}
\label{sub:DDD_simulation_results}

\begin{figure}
\centering
    \begin{subfigure}{0.49\textwidth}
    \centering
        \includegraphics[width=0.9\textwidth]{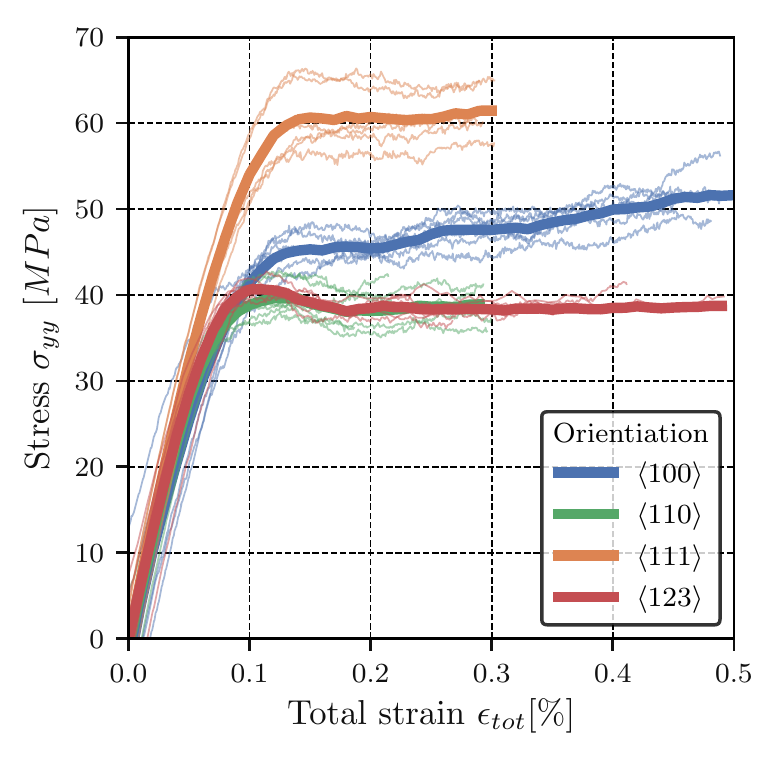}
        \caption{Evolution of the normal stress in tensile direction over the total strain.}
        \label{subfig:stress_strain}
    \end{subfigure}
    \hfill
    \begin{subfigure}{0.49\textwidth}
    \centering
        \includegraphics[width=0.9\textwidth]{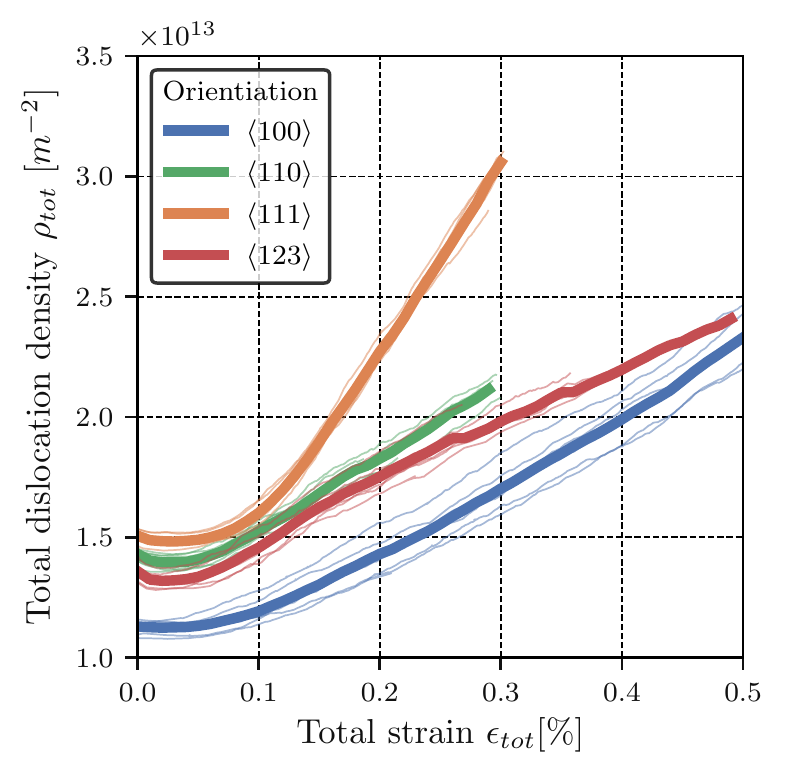}
        \caption{Evolution of the total dislocation density over the total strain.}
        \label{subfig:dens_strain}
    \end{subfigure}
    \caption{Evolution of the normal stress and the total dislocation density over the total strain in $\langle 100 \rangle$, $\langle 110 \rangle$, $\langle 111 \rangle$ and $\langle 123 \rangle$ orientations. Thin lines indicate single simulation and bold lines indicate the mean of the simulations for each orientation.
}
    \label{fig:stress_strain}
\end{figure}

The resulting stress-strain curves of all DDD simulations are shown in \autoref{subfig:stress_strain}.
Each single simulation has a jagged stress-strain curve during plastic deformation.
An orientation dependency for the stress-strain curves is observed as expected from the different Schmid factors for the high symmetry orientations, the initial yield strength in $\langle 111 \rangle$ orientation is around 60 MPa and therefore higher than in $\langle 100 \rangle$ with around 45 MPa or in $\langle 110 \rangle$ and in $\langle 123 \rangle$ with around 40 MPa.
We observe a significant hardening for the $\langle 100 \rangle$ orientation and a weak hardening for the $\langle 111 \rangle$ orientation.
No hardening is detected in the other orientations except a small stress drop after reaching the initial yield strength.

The evolution of the total dislocation density during the deformation is shown in \autoref{subfig:dens_strain}.
The different initial dislocation densities can be seen on the y-axis intercept, with the initial dislocation density deviating more than in $\langle 100 \rangle$.
Apart from a small dip, due to the activation of weak elements in the initial realaxed dislocation network, barley any change of the dislocation density is visible during the elastic regime.
During plastic deformation the dislocation density increases constantly for all orientations.
The dislocation density evolution in $\langle 111 \rangle$ shows the highest slope.
The dislocation density increase for the $\langle 100 \rangle$ orientation is slightly stronger than for the $\langle 110 \rangle$ and $\langle 123 \rangle$ orientations.

\begin{figure}
\centering
    \includegraphics[width=\textwidth]{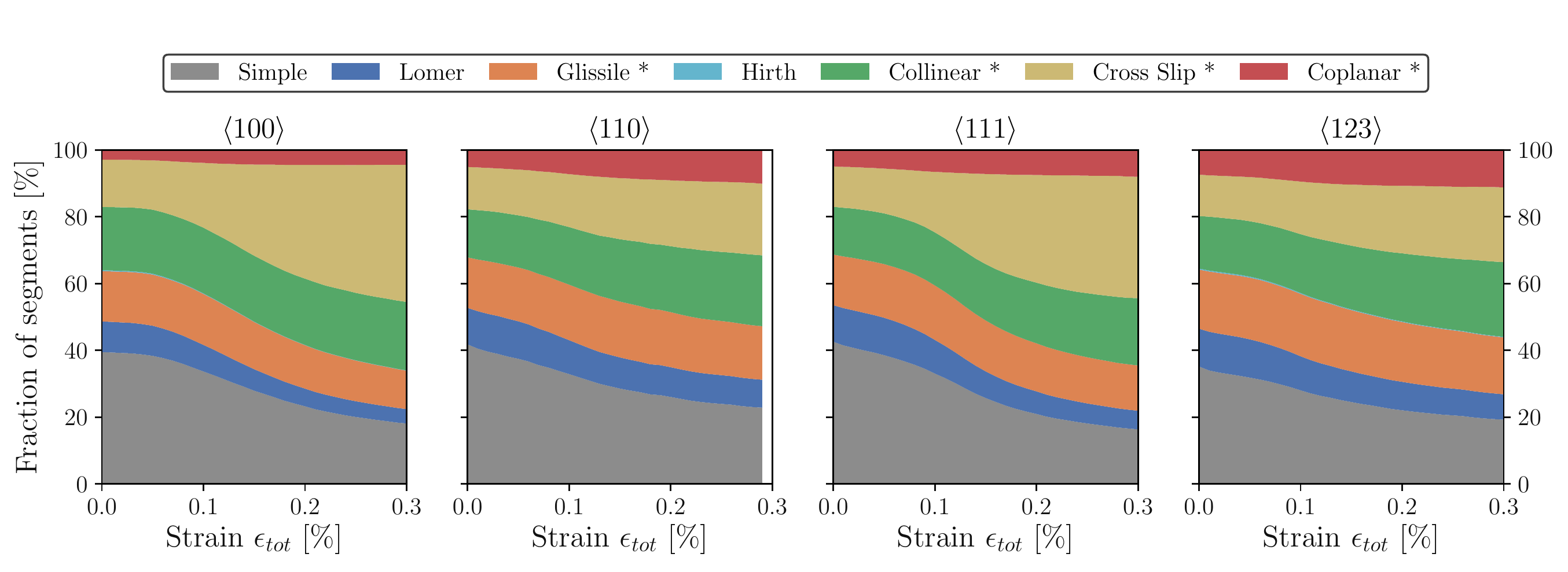}
    \caption{The average fraction evolution of mobile, so-called simple, dislocations and reactions with respect to all dislocations and reactions over the total strain $\epsilon_{tot}$ in the $\langle 100 \rangle$, $\langle 110 \rangle$, $\langle 111 \rangle$ and $\langle 123 \rangle$ orientation. 
    Simple dislocations as well as Lomer and Hirth reactions are physically existing junctions, whereas the reactions indicated with a (*) are virtual junctions. The Hirth reaction is very rare, which is why it is barely visible.
    }
    \label{fig:fraction_density_strain}
\end{figure}

The evolution of the fractions of "simple" glide sections of dislocation and sections involved in reactions are displayed in \autoref{fig:fraction_density_strain}.
For each category the total line length is calculated and normalized.
The results for each orientation are averaged over all six simulations.
The initial reaction configuration are very alike among the orientations and differs only slightly, e.g. the coplanar reaction is less present in $\langle 100 \rangle$ than in $\langle 123 \rangle$.
We observe a slight change of the fractions in the elastic regime and a more significant change during plastic deformation.
The fraction of coplanar, cross-slip and collinear reactions rises while straining whereas the fraction of Lomer and glissile reactions as well as of mobile, so-called simple, dislocations decreases.
The Hirth reaction has barely any contribution.
A major difference concerns the fraction of cross-slip, which rises more sharply in the plastic regime after a strain of 0.1\% in $\langle 100 \rangle$ and in $\langle 111 \rangle$ than in the other orientations.
The figure is cut at a strain of 0.3\% for better comparability among the orientations.
The evolution of the number of junctions and their average length is shown in \autoref{fig:junction_strain} of the Appendix.

\subsection{Physically based dislocation reaction rate models}
\label{sub:slip_system_interaction}

The coarse homogenization approach of the \textit{summed} reaction density evolution, see \autoref{eq: reaction denisity rate summed}, provides one constant coefficient $C_{react}$ for the behaviour of all interactions of each reaction type.
\autoref{tab: coefficient_summed} lists the reaction type dependent and orientation dependent results of the coefficients and the model quality $R^2$.
The model quality exceeds $0.42$ for all of the reaction rate predictions except for the Hirth reaction and reaches up to~$0.98$.
The model prediction performs better for the collinear reaction and the cross-slip mechanism than for the other reactions.
The reaction dependent model qualities in $\langle 111 \rangle$ orientation are above the model qualities in the other orientations, especially for glissle, Lomer and coplanar reactions.
The reaction rate coefficients are strongly reaction type dependent.
The coefficients of collinear reactions and of the cross-slip mechanisms are about one order of magnitude larger than glissile, coplanar and Lomer reactions and up to three orders of magnitude larger than the Hirth reaction.
Reactions with finite Burgers vector (physical lines) like Lomer and Hirth have a strong propensity to smaller reaction coefficients than reaction with zero net Burgers vector (virtual reactions).
Combining the results of the model qualities and the reaction coefficients, there is the tendency for a better model quality the higher the reaction coefficient. 

{
\renewcommand{\arraystretch}{1.1}
\begin{table}
  \centering
  \caption{Results of the coarse homogenization by using the \textit{summed} \autoref{eq: reaction denisity rate summed}, which provides one constant reaction coefficient for all interactions of each reaction type. The constant reaction coefficients $C_{react}$ and the model qualities $R^2$ are presented for $\langle 100 \rangle$, $\langle 110 \rangle$, $\langle 111 \rangle$ and $\langle 123 \rangle$ crystal orientation.}
    \begin{tabular}{p{0.15\textwidth} >{\centering}p{0.075\textwidth}  >{\centering}p{0.075\textwidth} >{\centering}p{0.075\textwidth} >{\centering}p{0.075\textwidth} >{\centering}p{0.075\textwidth} >{\centering}p{0.075\textwidth} >{\centering}p{0.075\textwidth} >{\centering\arraybackslash}p{0.075\textwidth}}
    \toprule[1pt]\midrule[0.3pt]
          & \multicolumn{2}{c}{$\mathbf{\langle 100 \rangle}$} & \multicolumn{2}{c}{$\mathbf{\langle 110 \rangle}$} & \multicolumn{2}{c}{$\mathbf{\langle 111 \rangle}$} & \multicolumn{2}{c}{$\mathbf{\langle 123 \rangle}$} \\
          & $C_{react}$ & $R^2$   & $C_{react}$ & $R^2$   & $C_{react}$ & $R^2$   & $C_{react}$ & $R^2$\\
    \hline
    Collinear & 0.5777 & 0.90  & 0.5488 & 0.86  & 0.6448 & 0.96  & 0.5933 & 0.82 \\
    Glissile & 0.0647 & 0.72  & 0.0823 & 0.53  & 0.1090 & 0.98  & 0.0937 & 0.56 \\
    Lomer & 0.0320 & 0.64  & 0.0552 & 0.54  & 0.0704 & 0.94  & 0.0544 & 0.63 \\
    Hirth & 0.0008 & 0.08  & 0.0001 & 0.04  & 0.0001 & 0.12  & 0.0009 & 0.01 \\
    Coplanar & 0.0635 & 0.59  & 0.1105 & 0.59  & 0.1377 & 0.96  & 0.1330 & 0.42 \\
    Cross Slip & 1.3920 & 0.88  & 0.6107 & 0.81  & 1.3138 & 0.97  & 0.6269 & 0.82 \\
    \midrule[0.3pt]\toprule[1pt]
    \end{tabular}
    \label{tab: coefficient_summed}
\end{table}
}

\begin{figure}
\centering
    \includegraphics[width=\textwidth]{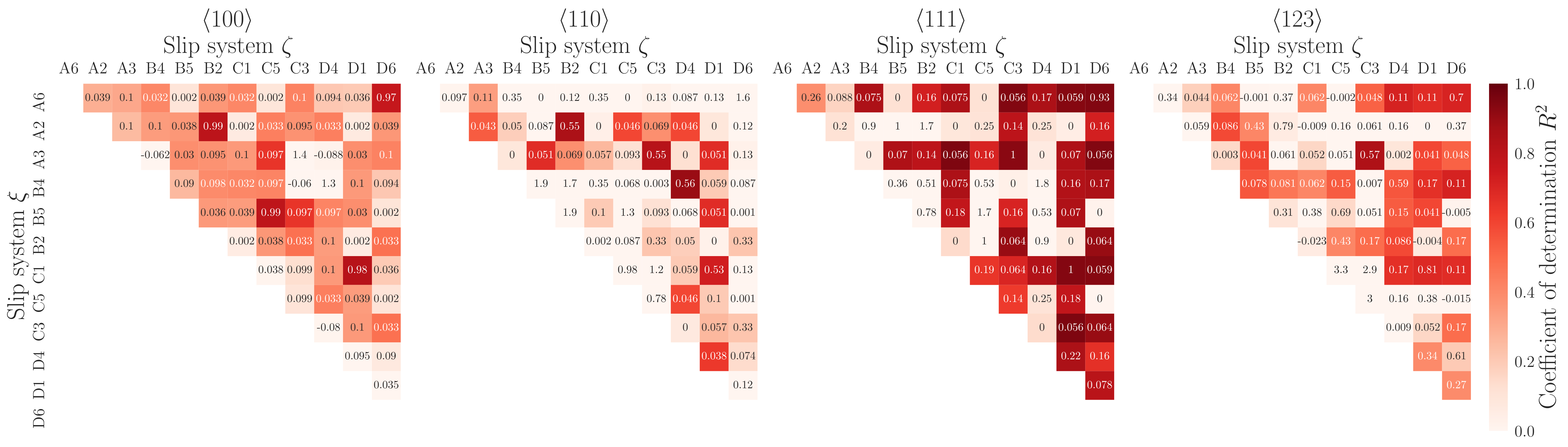}
    \caption{Slip system dependent heatmaps for $\langle 100 \rangle$, $\langle 110 \rangle$, $\langle 111 \rangle$ and $\langle 123 \rangle$ orientation by using the \textit{grouped} \autoref{eq: reaction denisity rate group}. The colored background indicates the model quality $R^2$, which reaches from negative values (here cut off at $0.0$) up to $1.0$ (see \autoref{subsub:regression_model}). Red squares indicate good model quality, white squares a bad model quality. The coefficients $C_{react}^{\xi,\zeta}$ of each interaction are printed inside the squares.
    }
    \label{fig:heatmap_r2}
\end{figure}

The \textit{grouped} approach leads to symmetric slip system dependent heatmaps, which are shown in \autoref{fig:heatmap_r2} for all considered orientations.
No obvious pattern is visible.
We observe a tendency that more reaction density rates are in accordance with our model for $\langle 100 \rangle$ orientation than for the other orientations, since darker read corresponds to a better model quality.
The model qualities show high values for most of the colored squares in $\langle 111 \rangle$ orientation.
Only a few slip system interactions are in accordance with our model in $\langle 110 \rangle$ and $\langle 123 \rangle$ orientation.

\begin{figure}
\centering
    \begin{subfigure}{0.48\textwidth}
        \includegraphics[width=\linewidth]{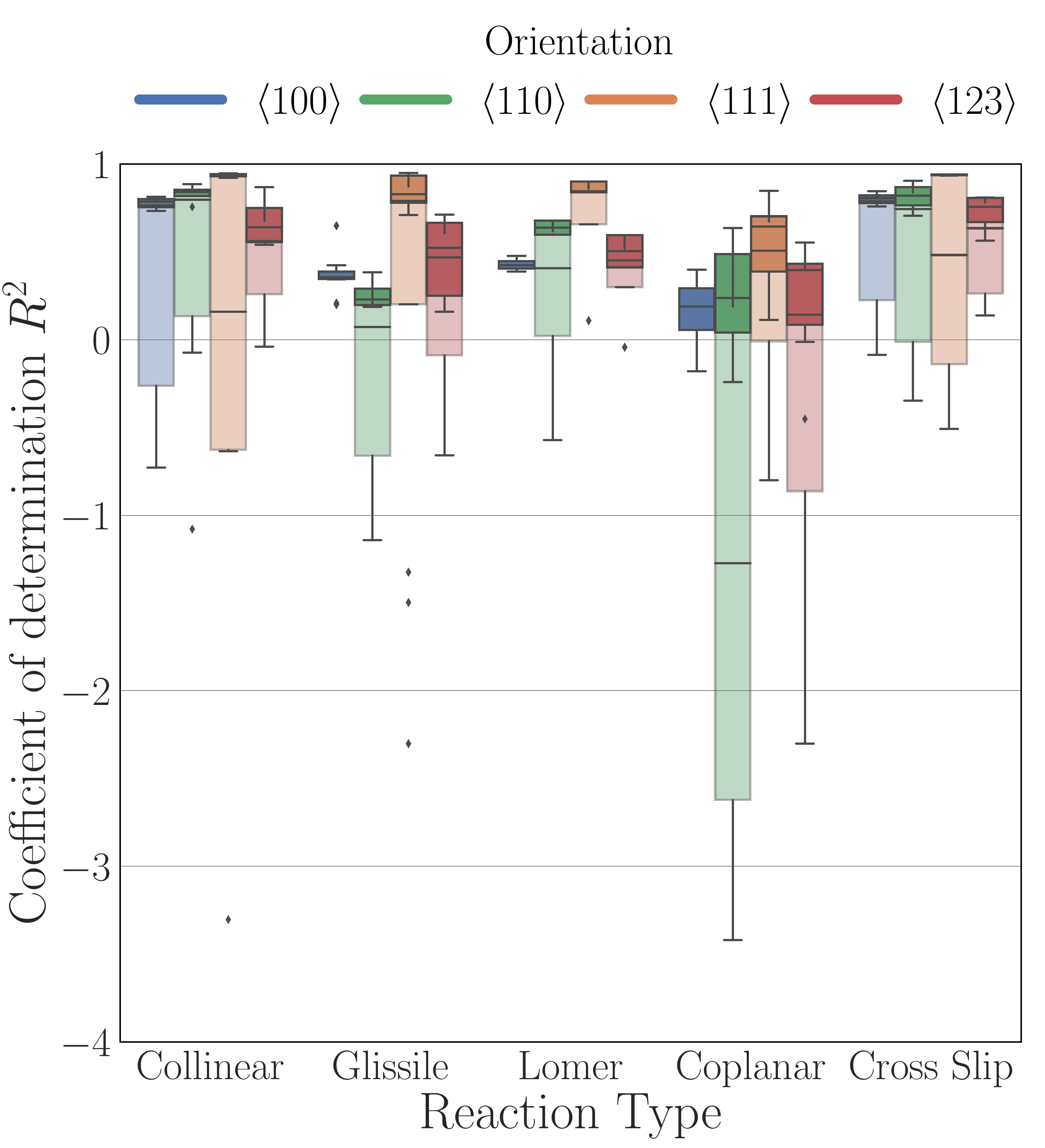}
        \caption{Coefficient of determination $R^2$}
        \label{subfig:boxplot_R2}
    \end{subfigure}
    \hspace*{\fill}
    \begin{subfigure}{0.48\textwidth}
        \includegraphics[width=\linewidth]{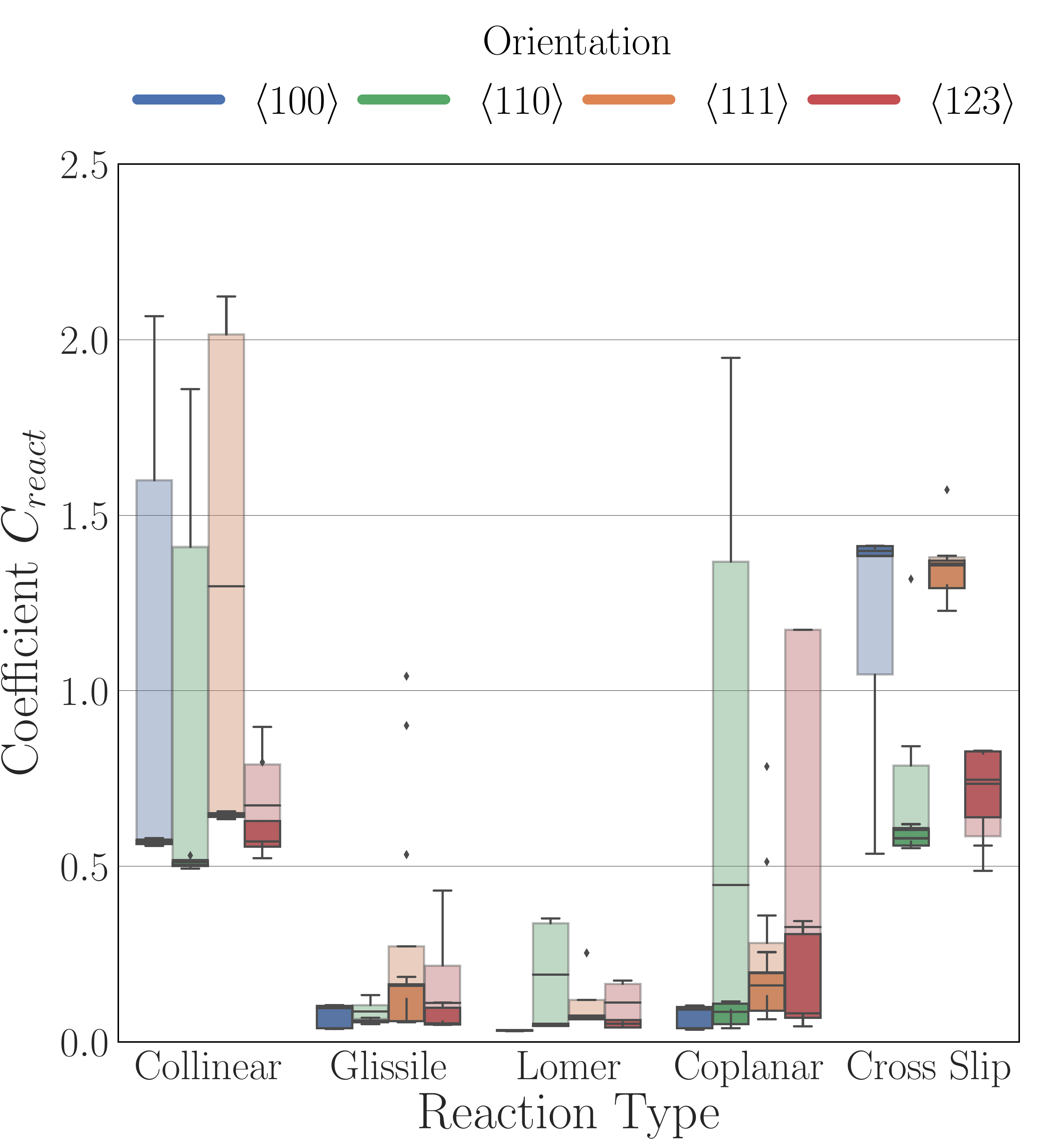}
        \caption{Coefficient $C_{react}$}
        \label{subfig:boxplot_C}
    \end{subfigure}
    \caption{Box plot of (\ref{subfig:boxplot_R2}) the model quality $R^2$ and of (\ref{subfig:boxplot_C}) the reaction coefficients $C_{react}^{\xi,\zeta}$ for each type of reaction and for each orientation. 
    The plots shown in pale color display the condensed results of the heatmaps in \autoref{fig:heatmap_r2}. 
    The plots shown in highly saturated color display the condensed results after the exclusion of the interactions between inactive slip systems (see \autoref{sub:Activity dependent model results}).
    The whiskers are defined at $1.5\%$ and $98.5\%$, the box defines the lower and the upper quartile and the median is shown by a thick line. The reduction of deviation is visible.
    }
    \label{fig:boxplots}
\end{figure}

We condense the information of the heatmaps in \autoref{fig:heatmap_r2} by aggregating each reaction type except for Hirth in each orientation.
The result is present in the low transparent (pale colored) box plots in \autoref{fig:boxplots}.
We observe a large deviation for the model qualities $R^2$ for each reaction type in \autoref{subfig:boxplot_R2}.
The Lomer reaction rate can be predicted best whereas our model fails for the other reaction types especially the coplanar reaction rate prediction.
An orientation dependence can not be identified.

The results for the coefficients $C_{react}^{\xi,\zeta}$ are shown in \autoref{subfig:boxplot_C}.
A large deviation of the coefficient values is visible for collinear and coplanar reaction as well as for the cross-slip mechanism.
The coefficients of glissile and Lomer reaction vary less than the coefficients of the other mechanisms.
The reaction coefficients are up to one order of magnitude smaller for glissile, Lomer and coplanar reactions than for the glissile reaction and cross-slip, which have both a net zero Burgers vector.
The value of the coefficients resembles in the different orientations except for the cross-slip mechanism, where the values are higher in $\langle 100 \rangle$ and $\langle 111 \rangle$ orientation than in the other orientations.

\subsection{Activity dependent model results}
\label{sub:Activity dependent model results}

In this section, the results of the \textit{grouped} approach of \autoref{eq: reaction denisity rate group}, which are presented in \autoref{sub:slip_system_interaction}, are subdivided by the activity of the interacting slip systems, i.e. into active-active, active-inactive and inactive-inactive interacting slip systems.
For collinear and coplanar reactions as well as for the cross-slip mechanism, this subdivision is applicable due to the consideration of only two interacting slip systems for each model equation.
In the \textit{grouped} approach of \autoref{eq: reaction denisity rate group} we considered groups of slip systems for the model equation for Lomer and glissile reactions, therefore the activity dependent subdivision for these reactions requires further declarations.
For the Lomer reaction densities, summed up within the respective slip system groups, the three respective slip systems are relevant in the \textit{grouped} approach. 
We consider the interaction as inactive-inactive, if none of the three slip systems is active, as active-inactive, if one of three slip systems is active, and as active-active, if two of the three slip systems are active.
E.g. the slip system group \{A6,C1,B4\} belongs to the inactive-inactive reactions in $\langle110\rangle$, while in $\langle100\rangle$ it belongs to the active-active reactions.
For the glissile reaction densities, two reaction pairs, each with two slip systems involved, are relevant for the model in the \textit{grouped} approach.
We choose the interaction as inactive-inactive, if both reaction pairs consist of at least one inactive slip system, as active-inactive, if only one of the two reaction pairs consists of two active slip systems, and as active-active, if all slip systems are active.

\begin{figure}
\centering
    \includegraphics[width=0.7\textwidth]{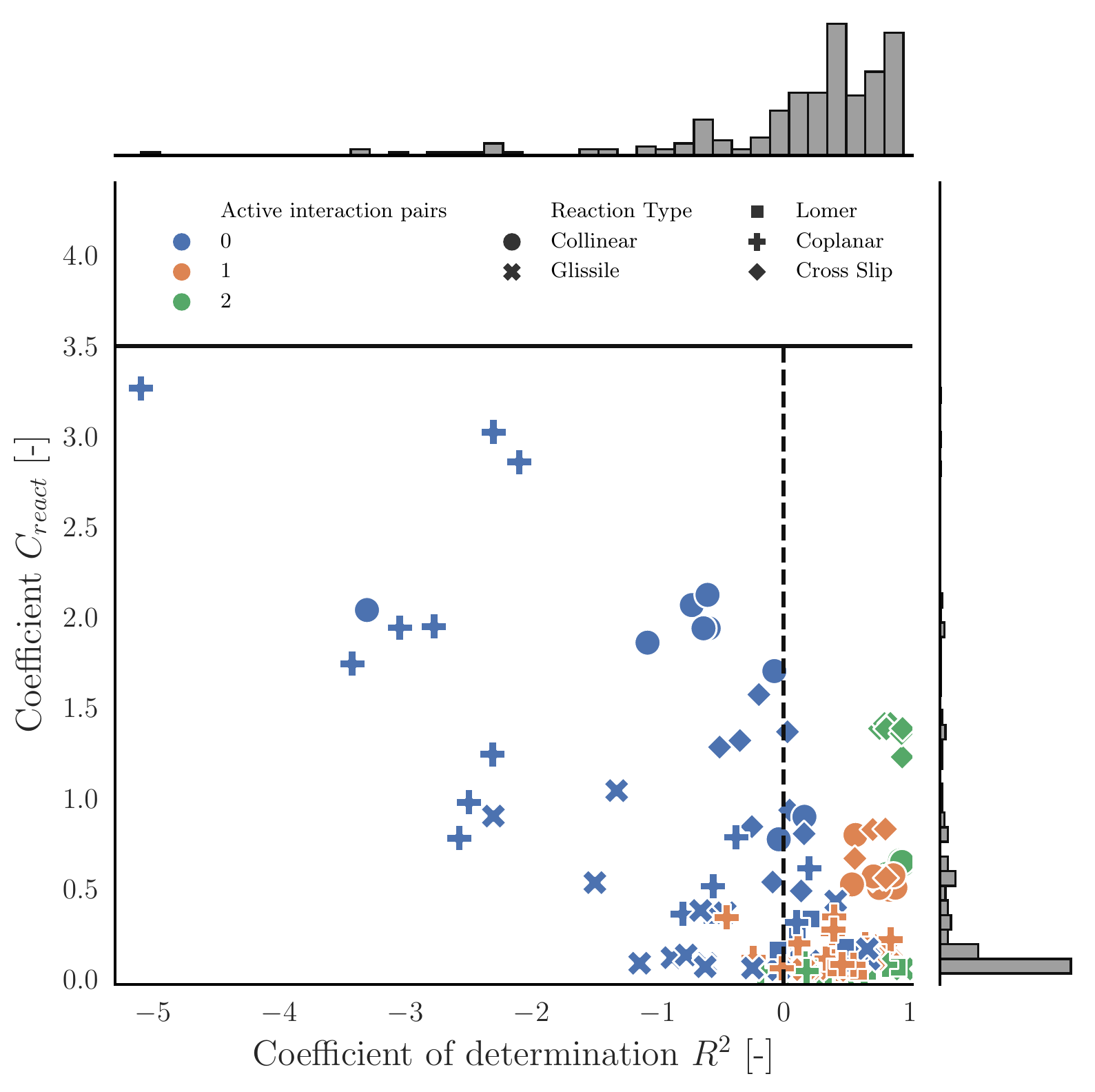}
    \caption{
    The results of the heatmaps in \autoref{fig:heatmap_r2} are subdivided into the interaction activity and into the reaction type, whereby the reaction coefficients $C_{react}^{\xi,\zeta}$ are plotted over the model quality $R^2$. Interaction pairs have 0 (inactive-inactive), 1 (active-inactive) or 2 (active-active) slip systems, which are considered as active. The distributions of the model qualities abd the reaction coefficients are shown on the top axis and on the right axis of the plot, respectively.
    }
    \label{fig:jointgrid_r2}
\end{figure}

The consideration of the slip system activity changes the results of the heatmap in \autoref{fig:heatmap_r2}.
The activity dependent result is shown in \autoref{fig:jointgrid_r2}.
We observe that the continuum reaction rate model is not able to predict the interaction between two inactive slip systems as can be seen by the bad model qualities $R^2$.
In contrast, the model can predict most of the interactions with at least one active slip systems.
If two slip systems are active, the model performs best.
There are some outliers for the inactive-inactive interactions, which have good model qualities as well, and for the active-inactive interactions, which seem to be not predictable.
The reaction coefficients $C_{react}^{\xi,\zeta}$ are subdivided by the activity dependent approach as well.
High reaction coefficients appear for inactive-inactive interactions, whereas active-inactive and active-active interactions have smaller reaction coefficients.
The reaction coefficients vary only slightly with at least one active slip system, whereas the reaction coefficients spread largely for the inactive-inactive interaction.

The subdivision by activity classifies the model into two groups, i.e., into the group with interactions between two inactive slip systems and into the group with interactions of at least one active slip system.
In the following, we observe more closely the group of interactions with at least one active slip system.

The exclusion of inactive-inactive slip system interactions leads to the highly saturated boxplots in \autoref{fig:boxplots}, whose quantitative results are listed in \autoref{tab:reaction_no_inactive}.
The mean model quality $R^2$ increases strongly and its deviation reduces significantly (see \autoref{subfig:boxplot_R2}).
The model qualities are greater than zero for all reaction types 
except for the coplanar reaction, where some model qualities are smaller than zero (see \autoref{fig:jointgrid_r2}).
The improvement of the mean model quality by only including slip systems with at least one active slip system is evident in contrast to the consideration of every slip system.
The deviation of the reaction coefficients $C_{react}^{\xi,\zeta}$ is strongly reduced and the coefficients tend to approach to specific values (see \autoref{subfig:boxplot_C}).
Small orientation dependent reaction coefficient differences are visible for all reaction types.
The activity dependent subdivision does not change the outcome, that the cross-slip mechanism has higher coefficients in $\langle 100 \rangle$ and in $\langle 111 \rangle$ orientation than in the other orientations with respect to the results of \autoref{sub:slip_system_interaction}. 
There is the peculiarity in $\langle 100 \rangle$ orientation, that there is no difference before and after the application of the activity dependent approach for the glissile, Lomer and coplanar reaction, which can explained by \autoref{tab: number_active_inactive} in the Appendix.
Due to the eight active slip systems in $\langle 100 \rangle$ orientation, there is no inactive-inactive slip system interaction for the glissile, Lomer and coplanar reactions.

As indicated in \autoref{tab:reaction_no_inactive}, the model show the highest $R^2$ values (correlating with the model quality) for the collinear reaction and the cross-slip mechanism, somewhat lower values for the Lomer reaction, and the lowest values for glissile and coplanar reactions.
Regarding the orientation, the model show the highest model quality in $\langle 111 \rangle$ orientation, where all reaction types except of the coplanar reaction have an $R^2 > 0.84$.

The mean coefficients $C_{react}^{\xi,\zeta}$ are reaction dependent, i.e., the coefficients of the reaction types can differ up to an order of magnitude, whereby the collinear reaction and the cross-slip mechanism have higher reaction coefficients than the other orientations.
The mean reaction coefficients $C_{react}^{\xi,\zeta}$ show an orientation dependence.
We observe the tendency of higher reaction coefficients in $\langle 111 \rangle$ orientation.
However, the reaction coefficients show not a clear trend for all reaction types in all orientations.
Thus, we detect a coupled dependency on reaction type and orientation.
We observe that the standard deviation of the reaction coefficients is small compared to the coefficients themselves.
This observation is even stronger for the collinear and Lomer reactions as well as for the cross-slip mechanism.
We observe that the standard deviation with respect to its coefficients is larger for the non-high symmetric $\langle 123 \rangle$ orientation than for the high symmetric orientations.
No orientation dependence is visible for the standard deviation in the high symmetric orientations.

{
\renewcommand{\arraystretch}{1.1}
\begin{table}
  \centering
  \caption{Result of the mean reaction coefficients, their standard deviation and the mean model performance of each reaction by using the \textit{grouped} approach of \autoref{eq: reaction denisity rate group} with exclusion of interactions between two inactive slip systems, i.e. at least one slip system is active for each interaction. The results are listed in $\langle 100 \rangle$, $\langle 110 \rangle$, $\langle 111 \rangle$ and $\langle 123 \rangle$ orientation for each type of reaction.}
    \begin{tabular}{l*{8}{c}}
    \toprule[1pt]\midrule[0.3pt]
    & \multicolumn{2}{c}{$\langle 100 \rangle$} & \multicolumn{2}{c}{$\langle 110 \rangle$} & \multicolumn{2}{c}{$\langle 111 \rangle$} &  \multicolumn{2}{c}{$\langle 123 \rangle$} \\
    & $C_{react}$ & $R^2$   & $C_{react}$ & $R^2$   & $C_{react}$ & $R^2$   & $C_{react}$ & $R^2$\\
    \hline
    Collinear & 0.569 $\pm$ 0.009 & 0.78  & 0.508 $\pm$ 0.016 & 0.83  & 0.646 $\pm$ 0.011 & 0.94  & 0.615 $\pm$ 0.123 & 0.67 \\
    Glissile & 0.079 $\pm$ 0.030 & 0.37  & 0.059 $\pm$ 0.007 & 0.26  & 0.129 $\pm$ 0.053 & 0.84  & 0.071 $\pm$ 0.032 & 0.46 \\
    Lomer & 0.032 $\pm$ 0.001 & 0.43  & 0.049 $\pm$ 0.003 & 0.64  & 0.070 $\pm$ 0.005 & 0.86  & 0.052 $\pm$ 0.012 & 0.50 \\
    Coplanar & 0.077 $\pm$ 0.030 & 0.16  & 0.080 $\pm$ 0.034 & 0.24  & 0.156 $\pm$ 0.067 & 0.55  & 0.174 $\pm$ 0.138 & 0.22 \\
    Cross Slip & 1.398 $\pm$ 0.017 & 0.80  & 0.583 $\pm$ 0.032 & 0.81  & 1.323 $\pm$ 0.084 & 0.94  & 0.720 $\pm$ 0.132 & 0.72 \\
    \midrule[0.3pt]\toprule[1pt]
    \end{tabular}
  \label{tab:reaction_no_inactive}
\end{table}
}

\subsection{Detailed model investigation of specific slip system interactions}
\label{sub:detailed_model_investigation}

We investigate the reaction rate model in detail by analyzing specific slip system interactions in a specific orientation.
As described in \autoref{sub:Activity dependent model results}, we classify the interactions into two groups, i.e. into a group with inactive slip systems and into a group with at least one active slip system.
We focus on the Lomer and the glissile reaction, since the reactions consist of three and four slip systems, respectively, due to the \textit{grouped} approach. 
We investigate the data in $\langle 111 \rangle$ orientation due to the peculiarity for the glissile reaction, that there is the case of interactions of four active slip systems as well as four inactive slip systems.

\begin{figure}
\centering
    \begin{subfigure}{0.48\textwidth}
        \includegraphics[width=0.9\textwidth]{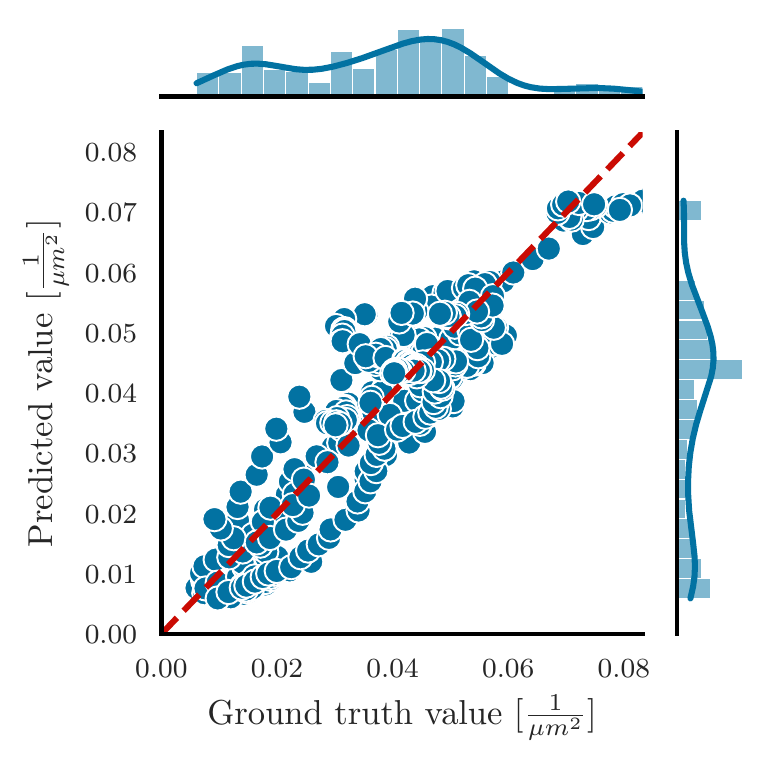}
        \caption{Active slip system interactions for the Lomer reaction: The involved slip systems are A6, B4 and C1. The prediction model score is $R^2 = 0.847$.}
        \label{subfig:specific_Lomer_active}
    \end{subfigure}
    \hspace*{\fill}
    \begin{subfigure}{0.48\textwidth}
        \includegraphics[width=0.9\textwidth]{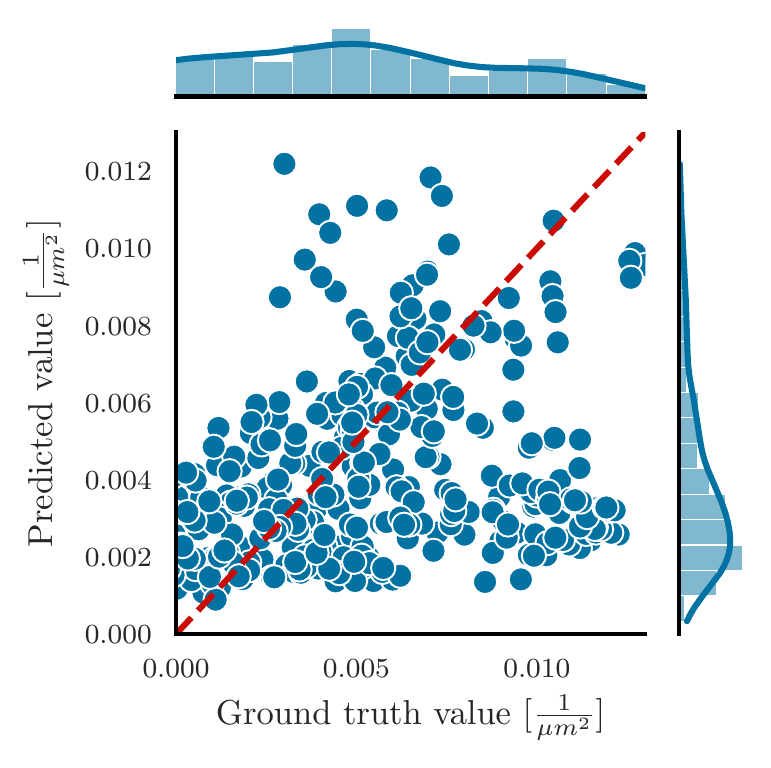}
        \caption{Inactive slip system interactions for the Lomer reaction: The involved slip systems are A2, C5 and D4. The prediction model score is $R^2 = 0.108$.}
        \label{subfig:specific_Lomer_inactive}
    \end{subfigure}
    \caption{Reaction density rate prediction of the \textit{grouped} model versus ground truth values (measured reaction density rate) of specific slip system reactions for the Lomer reaction in $\langle 111 \rangle$ orientation. The dashed red line indicates the perfect fit.
    }
    \label{fig:specific_Lomer}
\end{figure}

\begin{figure}
\centering
    \begin{subfigure}{0.48\textwidth}
        \includegraphics[width=0.9\textwidth]{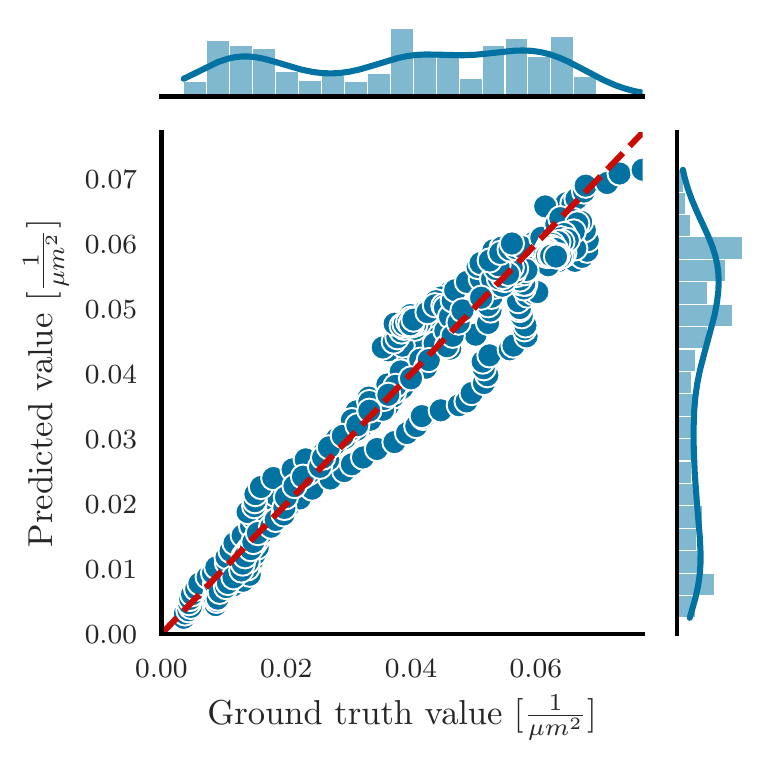}
        \caption{Active slip system interactions for the glissile reaction: The involved slip systems are A6, C3, A3 and D6 and create reaction length on slip system A2. The prediction model score is $R^2 = 0.935$.}
        \label{subfig:specific_Glissile_active}
    \end{subfigure}
    \hspace*{\fill}
    \begin{subfigure}{0.48\textwidth}
        \includegraphics[width=0.9\textwidth]{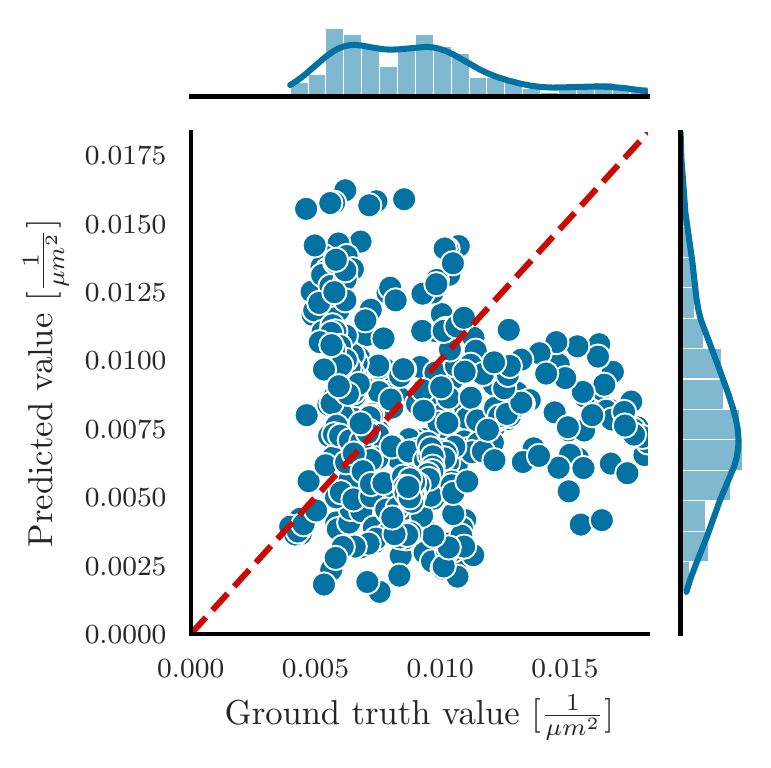}
        \caption{Inactive slip system interactions for the glissile reaction: The involved slip systems are A2, B5, B2 and C5 and create reaction length on slip system B4. The prediction model score is $R^2 = -1.324$.}
        \label{subfig:specific_Glissile_inactive}
    \end{subfigure}
    \caption{Reaction density rate prediction of the \textit{grouped} model versus ground truth values (measured reaction density rate) of specific slip system reactions for the glissile reaction in $\langle 111 \rangle$ orientation. The dashed red line indicates the perfect fit.}
    \label{fig:specific_Glissile}
\end{figure}

The prediction of the reaction rate of \autoref{eq: reaction denisity rate group} is plotted over the ground truth, which is equivalent to the measured reaction rate in our data sets.
We compare the model for the interaction of (a) active slip systems with the interaction for (b) inactive slip systems for the Lomer reaction in \autoref{fig:specific_Lomer} and for the glissile reaction in \autoref{fig:specific_Glissile}.
Additionally the distribution of the predicted and the ground truth values are plotted on the right and on the top axis, respectively.

\autoref{subfig:specific_Lomer_active} shows the model prediction for the Lomer reaction with two of the three involved slip systems active, i.e., for any combination between the active slip systems (A6 and C1) and the inactive slip system (B4) at least one slip system is active.
The model shows a good agreement with the data with a $R^2$ of 0.847. Both distributions have a peak at approximately the same value, which indicates a good agreement as well.
In \autoref{subfig:specific_Lomer_inactive} the interactions of the slip systems A2, C5 and D4 are used, which are all inactive.
We observe a scattering behaviour of our model, which means that the model is not able to predict the inactive interactions adequately.
The model quality of $R^2 = 0.108$ as well as the different distribution curves underline the poor model performance.

\autoref{subfig:specific_Glissile_active} shows the model prediction of the interactions between A6 and C3 and between A3 and D6, whereby all slip systems are active and whereby both interactions create glissile reaction density on the slip system A2. 
The model quality is close to the perfect fit with a $R^2$ of 0.935, although the distributions differ slightly between the predicted and the ground truth value.
In \autoref{subfig:specific_Glissile_inactive} the interactions between the inactive slip systems A2 and B5 and between B2 and C5 are investigated, which create glissile reaction on the slip system B4.
Our model shows a high scattering, which is confirmed by a $R^2$ of -1.324 
\footnote{Reminder: If $R^2 = 1$ the prediction is perfect, if $R^2 = 0$ the prediction is not better than a horizontal line and if $R^2 < 0$ the prediction is worse than a horizontal line.}
, which means that the performance is worse than a horizontal line. The distributions seem to be Gaussian distributions.

We observe for other slip system combinations that the model performs well in $\langle 111 \rangle$ for the Lomer and glissile reaction, the more slip systems are active.
The collinear reaction and the cross-slip mechanism show good model performance as well for the interaction of active slip systems, whereas the model shows a slight scattering for the coplanar reaction.
The model performs well in the other orientations, but slightly worse than in $\langle 111 \rangle$ orientation.

%CORRECTION DONE

\section{Discussion}
\label{sec:Discussion}

In this work, we present a data-driven analysis of dislocation networks by investigating fcc single crystals in $\langle 100 \rangle$, $\langle 110 \rangle$, $\langle 111 \rangle$ and $\langle 123 \rangle$ orientation under tensile loading.
We focus on rate equations for dislocation reactions, i.e. Lomer, Hirth, glissile, collinear and coplanar as well as the cross-slip mechanisms.
We use data scifence methods to determine constant, reaction type dependent and orientation-dependent coefficients, whereby the number of coefficients depends on the used reaction rate model.
We show that the reaction rate model can be improved by adding the information about the slip system activity. 
The investigation aims to evaluate physical-based formulations for dislocation network evolution based on DDD simulation data.
Furthermore, it aims to detect constant reaction coefficients for the characterization of the reaction kinetic as well as to transfer microstructural information from discrete to continuum modelling.

\subsection{Methodology}
\label{sub:dis_methodology}

Three dimensional dislocation networks have a complex topology, since dislocations interact strongly with each other and thereby form junctions.
To characterize the behaviour of the whole dislocation network during plastic deformation both data-driven methods for analysis due to the high complexity and to analyze the network in a homogenized volume-based approach for transferring information to higher scaled continuum based approaches are applied.

The application of data-driven methods fundamentally relies on the data quality.
Since the amount of experimental data on dislocation reaction details in 3d needed for data-driven investigations is not available so far, synthetic DDD data is used to surrogate this fact.
The comparability of the DDD data to experimental data has been investigated in literature~\cite{Kraft2010,el-awady_unravelling_2015,Akhondzadeh2020}.
Thus, a general acceptance of DDD data as database is
presumed~\cite{Bertin2020,Akhondzadeh2021}.
Qualitatively, experiments of tensile tested fcc single-crystals are in good accordance with DDD results considering stress-strain curves with respect to the yield stress of the crystal orientations, i.e. the yield stress is larger in $\langle 111 \rangle$ and $\langle 100 \rangle$ orientation compared to $\langle 110 \rangle$ and $\langle 123 \rangle$ orientation \cite{Luecke1952,Hosford_1960,takeuchi_work_1975,Franciosi_1982b}.
However, the investigated DDD data shows slightly lower hardening compared to the the aforementioned experiments (see \autoref{subfig:stress_strain}).
A reason might be the formulation of the cross-slip mechanism within the DDD framework.
This mechanism has a major impact on the hardening rate as discussed in \cite{Akhondzadeh2020} or may even lower the overall hardening rate as shown in \cite{Weygand2005}.
Nevertheless, the qualitative comparability of the DDD results with experiments is given and thus, the quality of the data base considered in the present analysis is considered to be sufficient to ensure a physically significant data-driven analysis.

We observe a homogenization limit for the reaction rate equations in the considered cases, since the model fails for prediction of the \textit{individual} approach of \autoref{eq: reaction denisity rate individual}.
This problem origins from the fact that within the considered data sets the deposited reaction product can not be traced back to a specific slip system dependent interaction for the Lomer and the glissile reaction.
Therefore, we introduce the \textit{grouped} approach of \autoref{eq: reaction denisity rate group}.
Thus, the homogenization limit for the continuum modelling of reaction rates starts at the scale, at which the backtracing of the reaction products is no longer possible.
Former analysis investigated each reaction mechanisms isolated \cite{Devincre2006,Kubin2008,Madec2008}.
Recent research focuses more on the dislocation network and its evolution \cite{sills_dislocation_2018,Sudmanns_2020,Akhondzadeh2020,Akhondzadeh2021}.
This work links the approaches of investigating dislocation networks with generating reaction coefficients for each type of mechanism.
In contrast to the isolated investigations, a three-dimensional dislocation network behaves more complex and therefore our goal of finding reaction coefficients of individual reactions within the network is not trivial.

Including the slip system activity complements our theory of network evolution.
The estimation of the slip system activities using the respective Schmid factors calculated a priori seems to be sufficiently accurate (compare \autoref{tab:Schmidfactor} and \autoref{fig:shear_strain}).
The applicability of the current model is limited to interactions with at least one active slip system.
The role of the reaction densities of inactive slip systems in the overall microstructure evolution is uncertain.
It is observed that active rather than inactive interacting slip systems are primarily the drivers of the newly evolved reaction length (compare the absolute values of the interaction of active and inactive slip systems in \autoref{fig:specific_Lomer} and \autoref{fig:specific_Glissile} in (a) and (b), respectively).
However, the dislocations on the inactive slip systems also serve as necessary reactants for the dislocations on the active slip systems.
It should be noted that dislocations on inactive slip systems can also be formed by the reaction product of active slip systems.
Further studies are needed to clarify whether and under which circumstances the interaction of inactive slip systems might be neglected.

Two important questions arise with respect to the reaction coefficients for the reaction rate equations: (I) What do the coefficients tell about network evolution and (II) how the coefficients themselves need to be interpreted in context of continuum modelling.
For (I) we show a qualitative difference between the junctions of this system and we assume that the absolute values of the coefficients are meaningful for the network evolution due to the good model qualities for the 24 DDD data sets including four different orientations.
However, a good model prediction in the considered cases does not necessarily mean that the model is physically profound.
Therefore, we need to confirm the model with further data in future studies.
For (II), we assume that the Lomer reaction coefficient can be interpreted easily, since it represent an non zero Burgers vector dislocation segment in DDD, i.e., the higher the coefficient, the higher the impact on the evolution of Lomer density.
Therefore, we assume that this coefficient value can be transferred directly to continuum approaches.
The other mechanisms (except the Hirth reaction) consist of virtual segments in our DDD dislocation structure.
Thus, the correlation of the change in virtual segment lengths with the newly formed physical dislocation lengths on the reaction slip system has to be investigated.
Nevertheless, we assume the transferability in continuum approaches for topological reasons, e.g., in the DDD dataset, a dislocation segment newly generated by a glissile reaction is described by both a physical dislocation line from the start to the end point and a virtual dislocation line closing the loop on the slip system.

In this work, the dislocation network is considered as a black box format. 
The input variables are acquainted and the discrete state of the network after a certain straining can be measured, but how the network behaves as a whole is physically not clear.
Therefore, we use a hybrid approach by combining theoretical considerations, which lead to physically based rate equations derived from domain knowledge, with machine learning tools, i.e. with linear regression.
Our data-driven tool should be as transparent as possible, since we want to proof our physics-based theory without any tool dependent artifacts.
Models with clear transparency are so-called white-box models.
This hybrid approach, a so called grey-box model, is crucial for the data-driven analysis of physical-based dislocation modelling, in order to validate physically based theories.
The results of this work show that this method is suitable for the theory of dislocation network evolution.
Limits of the theory are demonstrated, e.g., the activity dependent approach is discussed in \autoref{sub:dis_activity} and the limit of homogenization scale as well as the limit of model performance for certain reactions are discussed in \autoref{sub:dis_coeff}.
The model proposed in this work might be oversimplified, but based on this work's methodology other linear or nonlinear physically based models derived by domain knowledge can be verified.

\subsection{Slip system activity}
\label{sub:dis_activity}

We showed that the generation of new reaction density strongly depends on the activity of the interaction between individual slip systems. 
Based on our model results (see \autoref{sub:Activity dependent model results} and \autoref{sub:detailed_model_investigation}), we can clearly point out that (I) for each interaction mechanism, the slip system activity and therefore the amount of swept area by dislocations of each slip system controls most of the quantitative amount of the generated reaction density and that (II) the newly generated reaction density by slip system interactions is not tied to the active slip systems, but can be deposited on inactive slip systems as well. 
This mechanism is observed strongly for the glissile and the coplanar reaction.
\autoref{subfig:specific_Lomer_active} and \autoref{subfig:specific_Glissile_active} show a strong correlation with our \textit{grouped} model for the Lomer and the glissile reaction, respectively, and we observe this strong correlation for the other reactions as well. 
The specific glissile reaction in $\langle 111 \rangle$ orientation between $A6$ and $C3$ and between $A3$ and $D6$ deposit reaction density on $A2$, which is an inactive slip system, where no plastic shear occurs. The increase of dislocation density on inactive slip systems was formerly reported in \cite{Weygand2014,Stricker2015}, where it was contributed to the glissile reaction as well.

We choose the activity dependent model as inactive, when there is at least one inactive slip system for each of the two individual glissile reactions.
This might explain the blue cross marker outliers showing a good prediction on the right bottom in \autoref{fig:jointgrid_r2}.
Since both glissile reactions can occur between an inactive and an active slip system, the interaction is counted as inactive in total, although the active slip system contributes to the increase of reaction density.
In general, when inactive slip systems interact with each other, barely any reaction density results. 
However, the coefficients for the interaction of inactive slip systems are way higher compared to interactions with at least one active slip system.
We assume that this is on the one hand a scattering effect of DDD simulations and on the other hand an interaction, which our model can not depict.
\autoref{subfig:specific_Lomer_inactive} and \autoref{subfig:specific_Glissile_inactive} show that the model is not able to predict the inactive interactions.

The \textit{summed} approach is in accordance to the DDD data, thus, the overall influence of the inactive slip system interactions on the reaction densities can be considered as small, which is illustrated in \autoref{fig:App_pred_inactive_active} in the Appendix.
Comparing the quantitative values of the reaction density evolution, inactive and active interactions differ by an order of magnitude.
Like the glissile reaction, the coplanar reaction creates a new reaction density on a slip system, which does not belong to the two interacting slip systems and might belong to the inactive ones.
This mechanism is investigated in detail in \cite{Akhondzadeh2020}, where the coplanar reaction contributes additionally to the dislocation density rate. 
The fact that the Lomer reaction density is contributed by $50\%$ to each of the two interacting slip systems, can lead to a deposition of Lomer density on an inactive slip system as well, if only one of the two interacting slip systems is active.
An increase of reaction density on an inactive slip system is visible in \autoref{subfig:specific_Lomer_active}, where only two of the three slip systems are active.

In \autoref{sub:slip_system_interaction} we show that the \textit{summed} approach of \autoref{eq: reaction denisity rate summed} is applicable for all types of reaction, except for Hirth.
Due to the rare, singular occurrence of Hirth reactions, as seen in \autoref{fig:fraction_density_strain} and which was observed in \cite{Stricker_2018,Sudmanns_2020} as well, the continuum approach is not able to reproduce this reaction.
However, the other reactions, as well as the cross-slip mechanism, are in good agreement with the \textit{summed} model.
If sporadic reactions occur for which a reaction formation yields a very small gain in elastic energy, such as for Hirth reactions, the model cannot represent the evolution.
Stronger junction formation comes along with higher energy reduction. This is reflected in higher reaction coefficients in the homogenized network formulation.
The activity dependent approach is not applied to the \textit{summed} model, since we do not exclude any slip systems before we do the calculation of our model.
In future studies, a preselection of slip systems can be investigated.

The \textit{grouped} approach of \autoref{eq: reaction denisity rate group} displays a small scale of homogenization, which leads to a high scattered matrix of reaction coefficients and model qualities (see \autoref{fig:heatmap_r2}).
The classification by slip system activity shows the qualitative differences between the interactions and shows the cause of the high scattering, which is the inactive slip system interaction (see \autoref{fig:jointgrid_r2}).
Thus, the matrix is reduced by excluding the inactive interactions. \autoref{fig:boxplots} validates this approach that the slip system activity is important for a good model quality and for the network evolution in general. 
These results endorse the theory of continuum modelling with mobile and immobile dislocations.
This idea is incorporated in many continuum modelling approaches like in \cite{Ma2004,Ma2006,Fan2021,Roters2010}.
We recommend a combined modeling in continuum formulations with a dependence on mobility, i.e. a slip system activity, and on the dislocation density.

Regarding the high and non-high symmetry orientations, we observe an orientation dependence for the reaction coefficients for the \textit{summed} approach of \autoref{eq: reaction denisity rate summed} and for the activity dependent \textit{grouped} approach of \autoref{eq: reaction denisity rate group} (see \autoref{tab: coefficient_summed} and \autoref{tab:reaction_no_inactive}, respectively).
We assume that the number of active slip systems mainly contributes to the orientation dependence.
The number of active slip systems differ between four and eight for the multi-slip orientations.
The $\langle 123 \rangle$ orientation has four active slip systems, since we use a separation of a Schmid factor $S_{min}$ of $0.25$ for the activity (see \autoref{subsub:schmid_factor_dependent}).
However, there is clearly one main active slip system in $\langle 123 \rangle$, which is shown in \autoref{fig:shear_strain} in the Appendix, which challenges our chosen minimum Schmid factor. 
\autoref{tab: coefficient_summed} shows that the model quality performs slightly worse in orientations with only four active slip systems, i.e. in $\langle 110 \rangle$ and $\langle 123 \rangle$ for most of the reaction types.
However, we observe the best model performances for the $\langle 111 \rangle$ orientation with six active slip systems, although in $\langle 100 \rangle$ there are eight active slip systems.
This may arise from the fact, that the dislocation density evolution differs significantly in $\langle 111 \rangle$ orientation (see \autoref{subfig:dens_strain}).

The orientation dependency was observed in former computational interaction coefficient investigations for Franciosi yield stress investigations \cite{Kubin2008,Kubin2008a} and for dislocation multiplication rate investigations \cite{Stricker2015,Akhondzadeh2020} and agrees with our DDD data analysis. 
Future research is worthwhile to identify the origin of the orientation dependence for the dislocation network evolution, e.g. by a more detailed investigation of the dislocation microstructure and the investigation of further characteristic quantities for interactions.

The tendency of higher standard deviations of the coefficients in $\langle 123 \rangle$ orientation stays is contrast to the fact, that we use a minimum Schmid factor of $0.25$, where only four of the nine non-null active slip systems are considered.
The choice of only four active slip systems should lead to a greater reduction of the coefficient scattering.
However, due to the strong dominance of slip system $A3$ in the $\langle 123 \rangle$ orientation, the swept areas and therefore the shear strain rates of the other three active slip systems are comparably small (see \autoref{fig:shear_strain}).
For an appropriate homogenization, the chosen Schmid factor classification seems to be a good trade-off between number of included slip systems and loss of prediction quality.
For future studies we aim for an adaptation of the classification to a Schmid factor based function, in order to reduce the simplifications made by classifying into active and inactive.
We estimate that this function is of great importance for non-high symmetric orientations due to their heterogeneous distributed Schmid factors.

\subsection{Analysis of reaction coefficients}
\label{sub:dis_coeff}

This work aims to transfer the network evolution information of homogenized dislocation networks, e.g. reaction coefficients, from DDD to continuum approaches like CDD.
Our models correlate the network evolution rate by constant coefficients and by slip system dependent features, i.e. the plastic shear rate and the dislocation density.
Besides of the slip system activity, which is discussed in \autoref{sub:dis_activity}, the model quality and the coefficients differ by the applied equations, which depend on the scale of homogenization (\autoref{eq: reaction denisity rate individual}, \autoref{eq: reaction denisity rate summed} and \autoref{eq: reaction denisity rate group}) and on the specific type of reaction including the cross-slip mechanism.

The high evolution rates and therefore the high reaction coefficients of the cross-slip mechanism may origin from the fact that in $\langle 100 \rangle$ eight and in $\langle 111 \rangle$ six slip systems are active, whereas the other orientations have only four active slip systems.
As shown in \autoref{tab: number_active_inactive} in the Appendix, cross-slip occurs between two active slip systems for $\langle 100 \rangle$ and $\langle 111 \rangle$, while this is not the case for $\langle 110 \rangle$ and $\langle 123 \rangle$.
As in \cite{Stricker_2018} described, the cross-slip mechanism emerges new dislocation loops like the glissile mechanism.
It is expected to play a significant role in dislocation multiplication, which leads to an extension of continuum theory by a cross-slip term \cite{sudmanns_dislocation_2019}.

The findings in \cite{hussein_microstructurally_2015}, that cross-slip is increased at dislocation intersections, matches with our findings of increased cross-slip reaction for orientations with more active slip systems, which have a higher probability of an interaction of active slip systems.
\autoref{fig:junction_strain} shows the strong increase of the average junction length of cross-slip in $\langle 100 \rangle$ and $\langle 111 \rangle$ due to multiple cross-slip events, which is indicated by the increasing number of cross-slip junctions. 
Although the model has a high predictive quality for cross-slip in the considered data set, its physical profoundness, any validity limitations, and robustness must be determined for use in continuum formulations.
The comparison of the reaction coefficients needs to be done carefully, since rate equations of other researchers \cite{alankar_determination_2012, Akhondzadeh2020} do not include cross-slip.

The reaction coefficients differ in each orientation. We observe, that the number of active slip systems does not correlate directly with the quantitative value of the reaction coefficient. 
The range of the glissile reaction coefficients, which was estimated in \cite{Stricker_2018, Roters2019} by several orders of magnitude, could be reduced to less than one order of magnitude.
The coplanar reaction has an important role in the evolution of a dislocation network, especially for orientations other than $\langle 100 \rangle$, where the coplanar density increases stronger (see \autoref{fig:fraction_density_strain}).
The glissile as well as the coplanar reaction can lead to reaction density on inactive slip systems.
We assume that these two interactions are mainly responsible for the increase of dislocation density on inactive slip systems.
Since the reaction coefficient $C_{Coplanar}$ is similar to the glissile coefficient $C_{Glissile}$, we derive that both are equally strong effects for the increase of dislocation density on inactive slip systems.
The importance of the coplanar mechanism for continuum plasticity models is shown in \cite{Akhondzadeh2020}.
Our work underlines the importance due to findings that the kinetics of glissile and coplanar reaction seem to be similar in the considered data sets.

The collinear reaction coefficients are high compared to the other reactions and compared to former research \cite{sudmanns_dislocation_2019,Sudmanns_2020}.
It seems that the collinear mechanism is underestimated by about one order of magnitude.
Collinear reactions lead to a dislocation annihilation.
Therefore the high coefficient indicates a high annihilation rate.
However, the dislocation density increases steadily during plastic deformation (see \autoref{fig:initial dislocation density}).
This might origin from the following.
Since the collinear dislocation segments are virtual in DDD, the nodes of the segments can move apart during plastic deformation, so that the collinear reaction density increases steadily, although there might be no new collinear reaction. 
This could explain the high coefficients.
Before transferring the collinear coefficient from DDD to continuum modelling, the onward motion of a collinear segment after generationn needs to be clarified.

The Lomer reaction coefficient is in accordance with former research \cite{sudmanns_dislocation_2019,Sudmanns_2020} for the $\langle 100 \rangle$ orientation with $C_{Lomer} = 0.032$.
For the other orientations, the coefficient for the Lomer reaction is higher for both the \textit{summed} as well as the \textit{grouped} activity dependent approach. 
The scattering of the reaction density prediction curves might indicate that Lomer reactions are built and dissolved as well.
Lomer density does not increase on inactive groups but stay around the initial values (see Lomer in \autoref{fig:App_pred_inactive_active} in the Appendix).

We observe, that the coefficients differ from $0\%$ up to $31\%$ at most between the \textit{summed} and the activity dependent \textit{grouped} approach.
The tendency of slightly higher coefficient values for the \textit{grouped} approach stays in contrast to the fact, that inactive interacting slip systems are not included, which tend to have much higher values (see blue markers in \autoref{fig:jointgrid_r2}). 
However, nearly all of the reaction coefficient differences between the \textit{summed} and the \textit{grouped} approach remain within the standard deviation of the \textit{grouped} approach except for collinear and Lomer reaction in $\langle 110 \rangle$ orientation.

The model quality $R^2$ of the \textit{summed} approach outperforms the activity dependent \textit{grouped} approach for all reactions in all orientations except for the Lomer reaction in $\langle 110 \rangle$ orientation.
Thus, we derive a better homogenization for coarse homogenization than for a fine homogenization.
However, we show that the fine homogenized \textit{grouped} approach in combination with the slip system activity classification is able to predict the reaction rate with an acceptable model quality and a decent standard deviation of the reaction coefficient.

%CORRECTION DONE

\section{Conclusion}
\label{sec:Conclusion}

We introduce a dislocation density based model for the evolution of dislocation networks by considering different levels of homogenizing the slip system interaction.
This leads to three network evolution rate equations, which are specified by data-driven methods, i.e. multi-linear regression.
We validate the evolution equations for four different single crystal orientations, consisting of three high symmetric and one non-high symmetric orientation.

The analyses show that the reaction density evolution can be predicted well, if the homogenization level of the model does not exceed a certain threshold.
The prediction of every single interaction of all possible slip system combinations is not possible by the homogenized description.
However, groups of dislocation interactions can be predicted properly.
The proposed model couples the network evolution with a slip system activity dependent consideration.
The analysis shows a clear separation of the proposed model predictability between the inactive and active slip system interactions.
We observe an orientation dependency for each type of reaction for both the homogenized overall and the homogenized grouped consideration of each reaction.
In addition to the high symmetric orientations, the proposed model also shows accurate results for a non-high symmetric orientation.

The presented investigations suggest that a coupling of slip system interactions with the slip system activity is needed for the prediction of the network evolution.
The provided model uses physically based equations and provides constant coefficients for continuum modelling depending on the reaction type and slip system.
The number of coefficients differs depending on the need of the continuum model, i.e. it ranges from one to 24 coefficients depending on level of homogenization and the possible activity consideration.

%CORRECTION DONE

\section{Acknowledgement}
\label{sec:Acknowledgement}
 
The  financial  support  for  this work in the context of the DFG research project SCHU 3074/4-1  is gratefully acknowledged.

%CORRECTION DONE

\setcounter{figure}{0}
\counterwithin{figure}{section}
\setcounter{table}{0}
\counterwithin{table}{section}
\appendix
\label{sec:Appendix}

\section{Additional information of the DDD data sets}
\label{sec:Appendix_Additional}

\begin{figure}[H]
\centering
    \includegraphics[width=\textwidth]{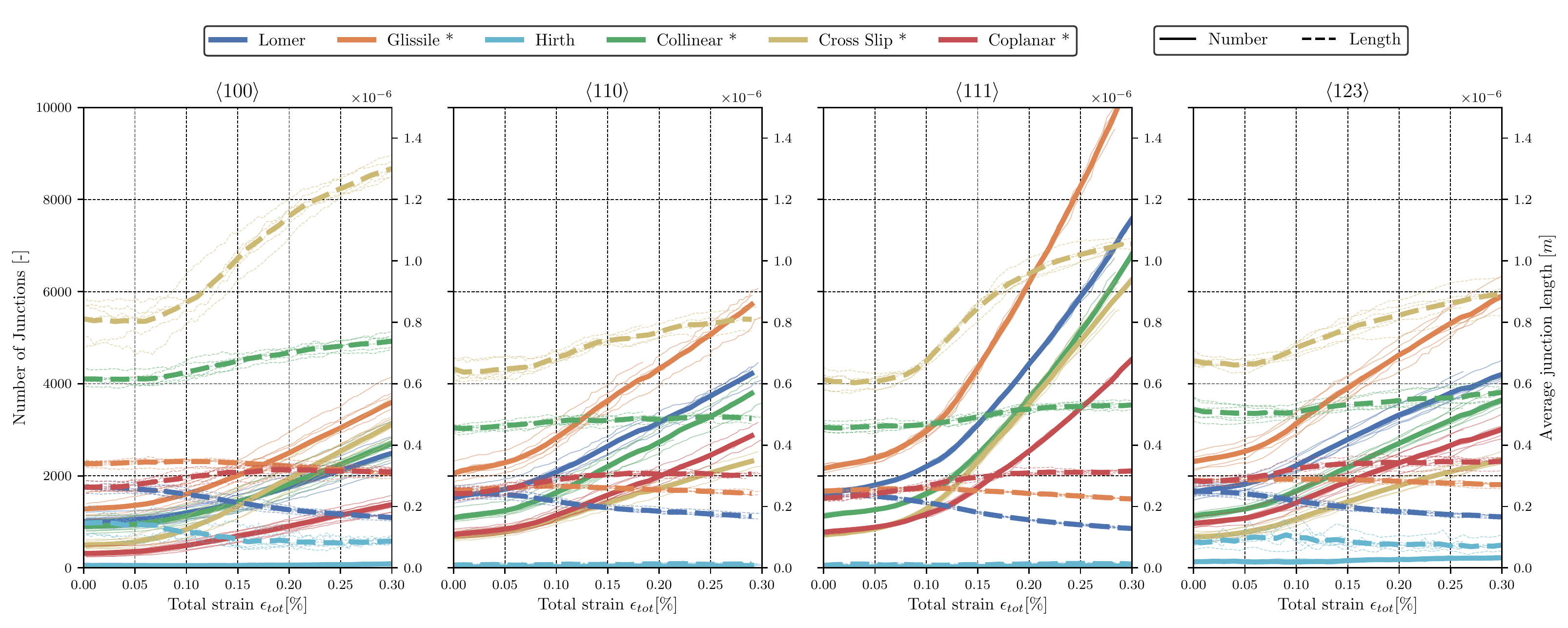}
    \caption{Evolution of the number of junctions and the average junction length for different reactions over the total strain in the orientations $\langle 100 \rangle$, $\langle 110 \rangle$, $\langle 111 \rangle$ and $\langle 123 \rangle$. Lomer and Hirth are physically existing junctions, whereas the reactions indicated with a (*) are virtual junctions.
    }
    \label{fig:junction_strain}
\end{figure}

\begin{figure}[H]
\centering
    \includegraphics[width=\textwidth]{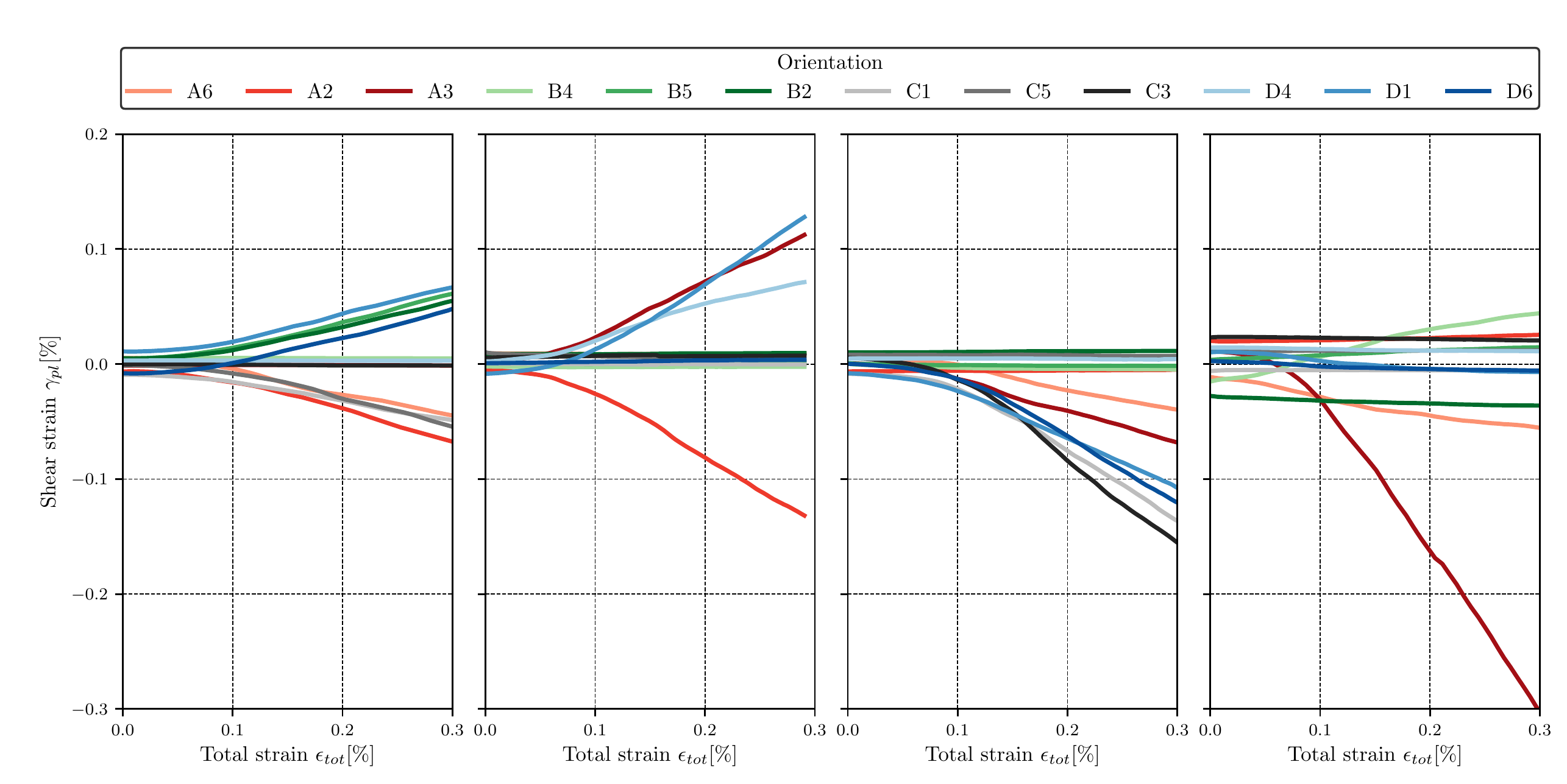}
    \caption{Plastic shear strain of each slip system for one dataset in each orientation $\langle 100 \rangle$, $\langle 110 \rangle$, $\langle 111 \rangle$ and $\langle 123 \rangle$.
    }
    \label{fig:shear_strain}
\end{figure}

In this Appendix we show additional results of the fcc single crystals under tensile loading.
\autoref{fig:junction_strain} displays the evolution of the number of junctions and the evolution of the average junction length over the total strain in $\langle 100 \rangle$, $\langle 110 \rangle$, $\langle 111 \rangle$ and $\langle 123 \rangle$ orientation.
We observe that the number of junctions increases while straining the crystals for all reactions and in all orientations.
The average junction length increases for the collinear and coplanar reaction and the cross-slip mechanism, whereas it decreases for the Lomer and glissile reactions.
\autoref{fig:shear_strain} shows the plastic shear strain evolution over the total strain in every investigated orientation.
Inactive slip systems are clearly visible, since they have a straight horizontal course.
Active slip systems shear plastically in the positive or negative direction.
The number of active slip systems differs with the orientation. An absolute quantitative difference for the plastic shear is apparent for the different orientations.
Especially in $\langle 110 \rangle$ and $\langle 123 \rangle$ orientations the plastic shear is dominated by a few slip systems.
We observe that the shear strain curves of the active slip systems vary slightly for the high symmetry orientations, although the Schmid factors are equal for these slip systems.
\autoref{tab: Schmid-Boas} shows the Schmid-Boas slip system notation.

{
\renewcommand{\arraystretch}{1.1}
\begin{table}[H]
    \centering
    \begin{tabular}{cccc cc cccccc}
    \toprule[1pt]\midrule[0.3pt]
         \multicolumn{4}{c}{Slip plane normal} &&& \multicolumn{6}{c}{Burgers vector}\\
         A&B&C&D &&& 1&2&3&4&5&6 \\
         $(1 \overline{1} \overline{1})$ & $(1 1 1)$ & $(\overline{1} \overline{1} 1)$ & $(\overline{1} 1 \overline{1})$ &&& $\frac{1}{2}[0 1 1]$  & $\frac{1}{2}[0 1 \overline{1}]$ & $\frac{1}{2}[1 0 1]$ & $\frac{1}{2}[\overline{1} 0 1]$ &  $\frac{1}{2}[\overline{1} 1 0]$ & $\frac{1}{2}[1 1 0]$\\
    \midrule[0.3pt]\toprule[1pt]
    \end{tabular}
    \caption{Schmid-Boas notation for the slip plane normal and the Burgers vector of fcc slip systems.}
    \label{tab: Schmid-Boas}
\end{table}
}

\section{Number of active and inactive slip systems involved in the dislocation reactions}
\label{ssec:NumberOfActivePairs}

\autoref{tab: number_active_inactive} shows additional information about the activity of an interaction between two slip systems in $\langle 100 \rangle$, $\langle 110 \rangle$, $\langle 111 \rangle$ and $\langle 123 \rangle$ orientation.
It complements the information about inactive-inactive (0), active-inactive (1) and active-active (2) slip system interactions, which are introduced in \autoref{sub:Activity dependent model results}.
We observe the peculiarity in $\langle 100 \rangle$ orientation, that there is no inactive-inactive interaction for the glissile, Lomer and coplanar reaction due to the activity of eight slip systems.
All the other reactions in every investigated orientations has an inactive-inactive as well as an interaction with at least one active slip system (active-inactive or active-active).
This binary classification is used in \autoref{sub:Activity dependent model results} and in \autoref{sub:detailed_model_investigation}.

{
\renewcommand{\arraystretch}{1.1}
\begin{table}[H]
  \centering
  \caption{Number of active and inactive slip systems involved in the binary dislocation reactions for the different crystal orientations (cp. \autoref{fig:jointgrid_r2}). Self-interaction indicates the number of active slip systems. According to the interaction matrix in \autoref{fig:interaction_matrix}, each orientation counts 6 collinear and cross slip reaction pairs, 12 Lomer, Hirth and coplanar reaction pairs, and 24 glissile reaction pairs.}
    \begin{tabular}{m{0.16\textwidth} >{\centering}p{0.035\textwidth}  >{\centering}p{0.035\textwidth} >{\centering}p{0.035\textwidth}
    >{\centering}p{0.001\textwidth}
    >{\centering}p{0.035\textwidth} >{\centering}p{0.035\textwidth} >{\centering}p{0.035\textwidth}
    >{\centering}p{0.001\textwidth}
    >{\centering}p{0.035\textwidth} >{\centering}p{0.035\textwidth} >{\centering}p{0.035\textwidth}
    >{\centering}p{0.001\textwidth}
    >{\centering}p{0.035\textwidth} >{\centering}p{0.035\textwidth} >{\centering\arraybackslash}p{0.04\textwidth}}
    \toprule[1pt]\midrule[0.3pt]
          & \multicolumn{3}{c}{$\mathbf{\langle 100 \rangle}$} && \multicolumn{3}{c}{$\mathbf{\langle 110 \rangle}$} && \multicolumn{3}{c}{$\mathbf{\langle 111 \rangle}$} && \multicolumn{3}{c}{$\mathbf{\langle 123 \rangle}$} \\
          Number of active slip systems & $2$  & $1$ & $0$   && $2$ & $1$ & $0$   && $2$ & $1$ & $0$   && $2$ & $1$ & $0$\\
    \hline
    Self-Interaction &8&0&4 &&4&0&8 &&6&0&6 &&4&0&8\\
    Collinear &4&0&2 &&0&4&2 &&3&0&3 &&0&4&2\\
    Glissile &8&16&0 &&0&16&8 &&6&12&6 &&2&12&10\\
    Lomer &4&8&0 &&2&4&6 &&3&6&3 &&2&4&6\\
    Hirth &8&0&4 &&2&4&6 &&0&12&0 &&1&6&5\\
    Coplanar &4&8&0 &&2&4&6 &&3&6&3 &&1&6&5\\
    Cross Slip &4&0&2 &&0&4&2 &&3&0&3 &&0&4&2\\
    \midrule[0.3pt]\toprule[1pt]
    \end{tabular}
    \label{tab: number_active_inactive}
\end{table}
}

\section{Extended results of the data-driven analysis}
\label{sec:Appendix_extendend_results}

\begin{figure}[H]
\centering
    \includegraphics[width=0.3\textwidth]{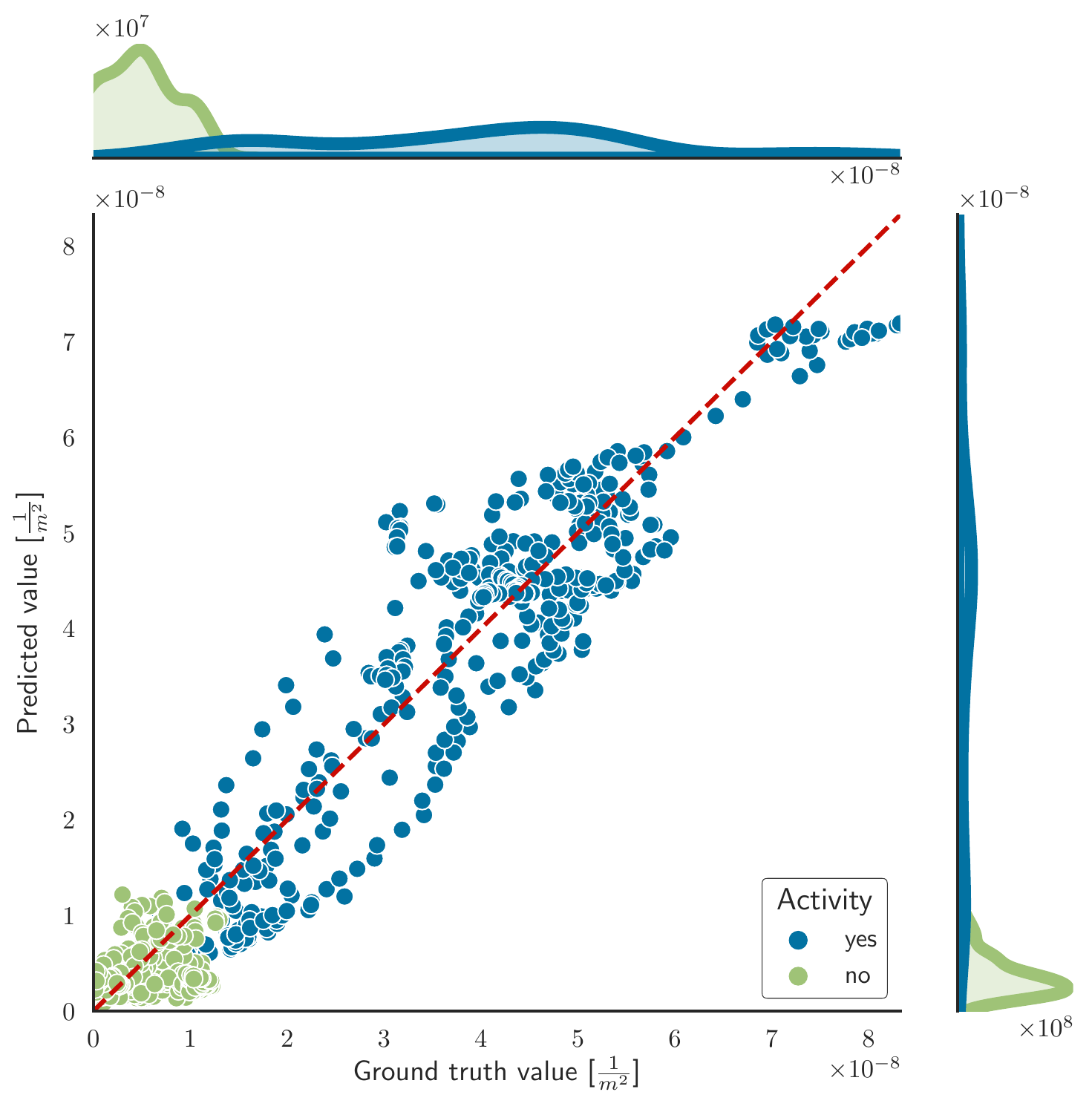}
    \includegraphics[width=0.3\textwidth]{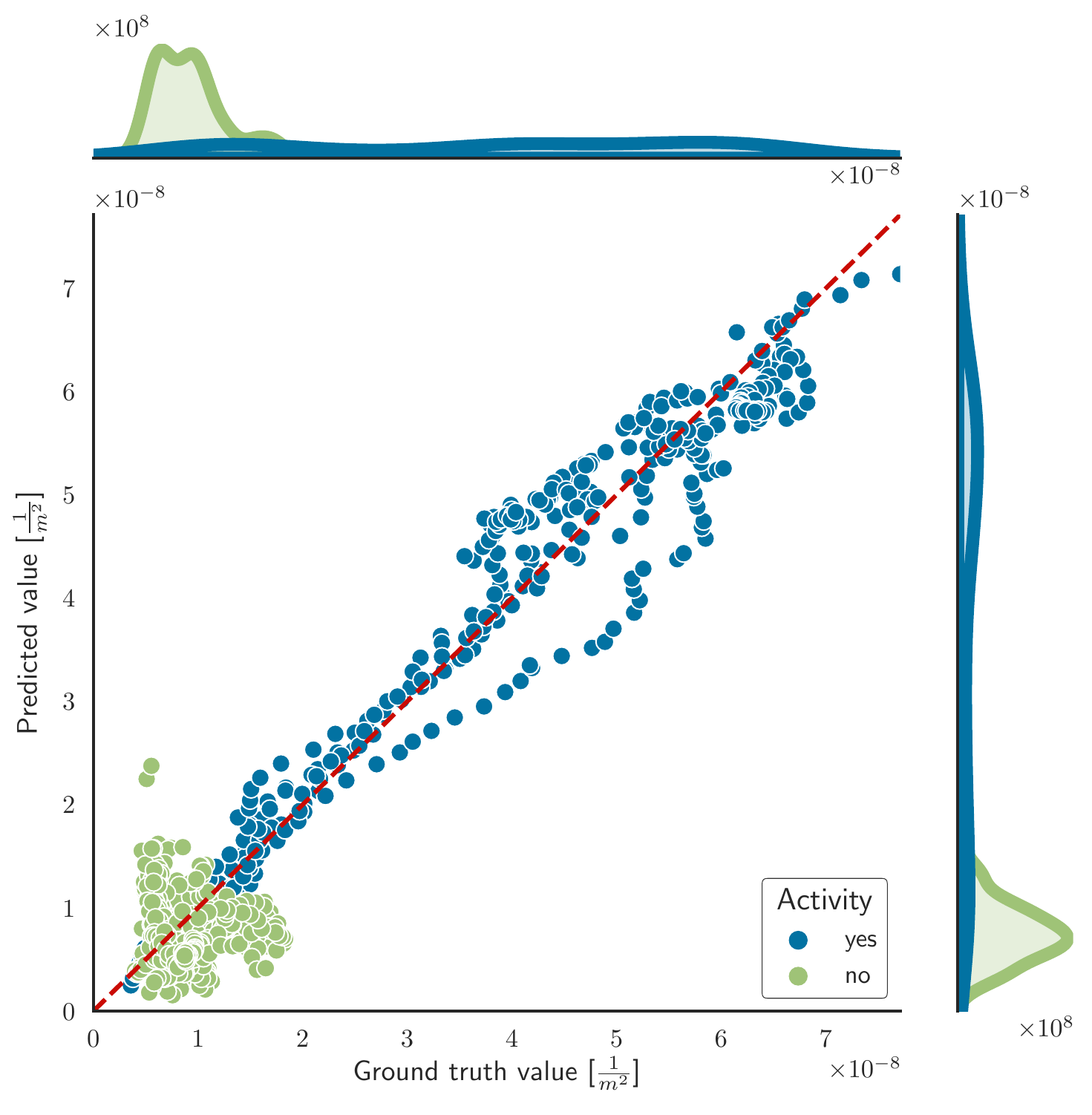}
    \includegraphics[width=0.3\textwidth]{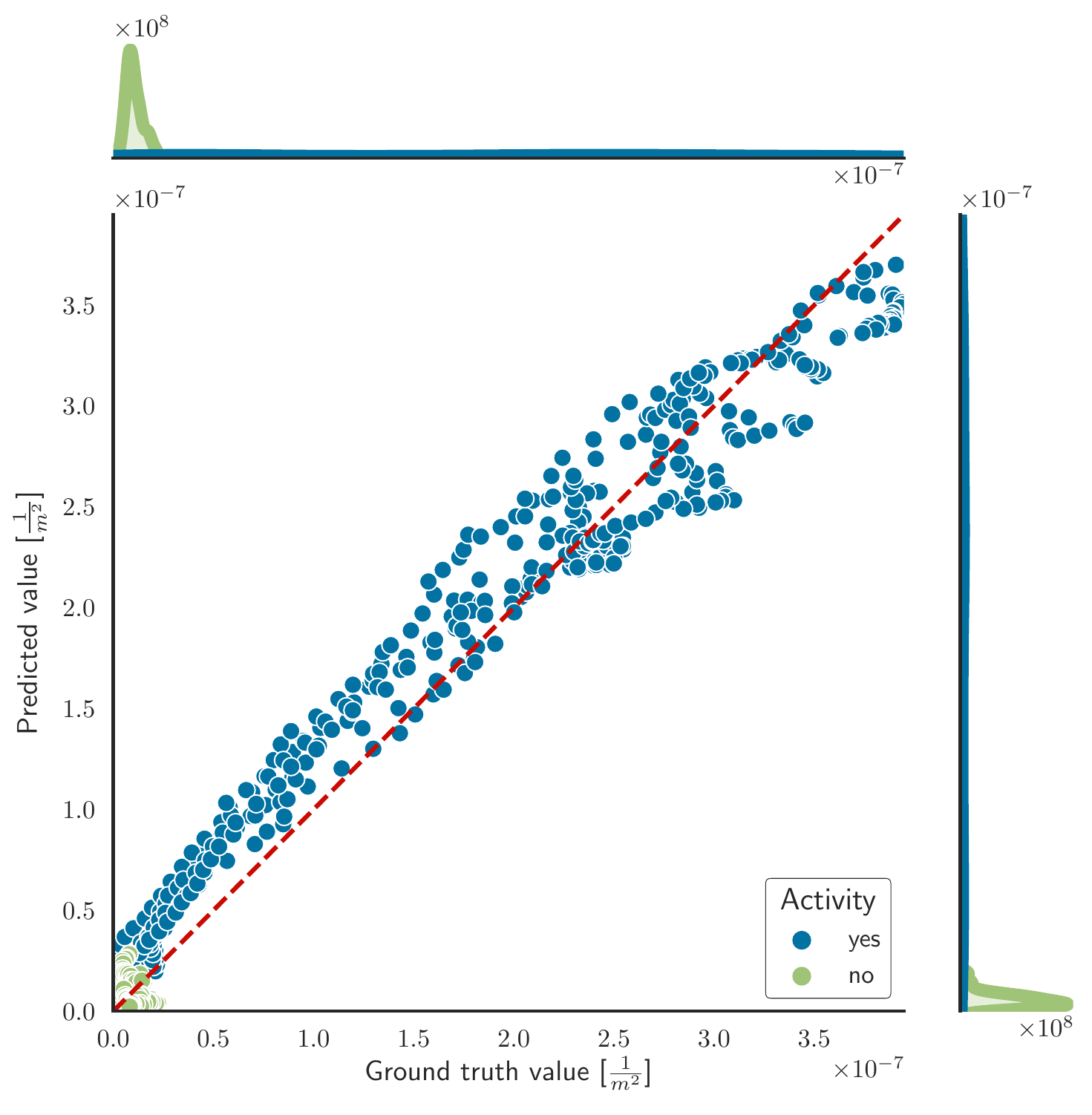}
    \caption{(Left) Lomer, (center) glissile and (right) collinear reaction rate prediction versus ground truth values (measured reaction density rate) of specific slip system reactions in $\langle 111 \rangle$  of the \textit{grouped} model approach.
    }
    \label{fig:App_pred_inactive_active}
\end{figure}

In this Appendix the results of the inactive and active slip system interaction in $\langle 111 \rangle$ in \autoref{sub:detailed_model_investigation} are combined in one figure for the Lomer and the glissile reaction.
\autoref{fig:App_pred_inactive_active} shows additionally the interaction of two inactive slip systems and two active slip systems for the collinear reaction.
We observe that the contribution of the interactions of inactive slip systems is an order of magnitude smaller than between active slip systems.
Thus, the distribution peaks at the small prediction and ground truth values for inactive interactions, whereas the distribution of active slip systems are equally distributed.

%CORRECTION DONE

%% The Appendices part is started with the command \appendix;
%% appendix sections are then done as normal sections
%% \appendix

%% \section{}
%% \label{}

%% If you have bibdatabase file and want bibtex to generate the
%% bibitems, please use
%%

\bibliographystyle{elsarticle-num.bst}
%%  \bibliography{<your bibdatabase>}

%% else use the following coding to input the bibitems directly in the
%% TeX file.

\bibliography{DataAnalysis}

\providecommand{\noopsort}[1]{}\providecommand{\singleletter}[1]{#1}%
\begin{thebibliography}{10}
\expandafter\ifx\csname url\endcsname\relax
  \def\url#1{\texttt{#1}}\fi
\expandafter\ifx\csname urlprefix\endcsname\relax\def\urlprefix{URL }\fi
\expandafter\ifx\csname href\endcsname\relax
  \def\href#1#2{#2} \def\path#1{#1}\fi

\bibitem{livingston_density_1962}
J.~D. Livingston,
  \href{http://www.sciencedirect.com/science/article/pii/0001616062901207}{The
  density and distribution of dislocations in deformed copper crystals}, Acta
  Metallurgica 10~(3) (1962) 229--239.
\newblock \href {https://doi.org/10.1016/0001-6160(62)90120-7}
  {\path{doi:10.1016/0001-6160(62)90120-7}}.
\newline\urlprefix\url{http://www.sciencedirect.com/science/article/pii/0001616062901207}

\bibitem{basinski_dislocation_1964}
Z.~S. Basinski, S.~J. Basinski,
  \href{https://doi.org/10.1080/14786436408217474}{Dislocation distributions in
  deformed copper single crystals}, The Philosophical Magazine: A Journal of
  Theoretical Experimental and Applied Physics 9~(97) (1964) 51--80.
\newblock \href {https://doi.org/10.1080/14786436408217474}
  {\path{doi:10.1080/14786436408217474}}.
\newline\urlprefix\url{https://doi.org/10.1080/14786436408217474}

\bibitem{pande_dislocation_1971}
C.~S. Pande, P.~M. Hazzledine,
  \href{https://doi.org/10.1080/14786437108217068}{Dislocation arrays in
  {Cu}-{Al} alloys. {I}}, The Philosophical Magazine: A Journal of Theoretical
  Experimental and Applied Physics 24~(191) (1971) 1039--1057.
\newblock \href {https://doi.org/10.1080/14786437108217068}
  {\path{doi:10.1080/14786437108217068}}.
\newline\urlprefix\url{https://doi.org/10.1080/14786437108217068}

\bibitem{Kubin2008a}
L.~Kubin, B.~Devincre, T.~Hoc, Toward a physical model for strain hardening in
  fcc crystals, Materials Science and Engineering: A 483-484 (2008) 19--24.
\newblock \href {https://doi.org/10.1016/j.msea.2007.01.167}
  {\path{doi:10.1016/j.msea.2007.01.167}}.

\bibitem{Andreoni_2020}
W.~Andreoni, S.~Yip (Eds.),
  \href{http://swbplus.bsz-bw.de/bsz1694052567cov.htmhttps://doi.org/10.1007/978-3-319-44677-6}{Handbook
  of Materials Modeling : Methods: Theory and Modeling}, 2nd Edition, Springer
  eBook Collection, Springer International Publishing, Cham, 2020.
\newline\urlprefix\url{http://swbplus.bsz-bw.de/bsz1694052567cov.htmhttps://doi.org/10.1007/978-3-319-44677-6}

\bibitem{Roters2019}
F.~Roters, M.~Diehl, P.~Shanthraj, P.~Eisenlohr, C.~Reuber, S.~Wong, T.~Maiti,
  A.~Ebrahimi, T.~Hochrainer, H.-O. Fabritius, S.~Nikolov, M.~Fri{\'{a}}k,
  N.~Fujita, N.~Grilli, K.~Janssens, N.~Jia, P.~Kok, D.~Ma, F.~Meier,
  E.~Werner, M.~Stricker, D.~Weygand, D.~Raabe, {DAMASK} {\textendash} the
  düsseldorf advanced material simulation kit for modeling multi-physics
  crystal plasticity, thermal, and damage phenomena from the single crystal up
  to the component scale, Computational Materials Science 158 (2019) 420--478.
\newblock \href {https://doi.org/10.1016/j.commatsci.2018.04.030}
  {\path{doi:10.1016/j.commatsci.2018.04.030}}.

\bibitem{Kocks2003}
U.~Kocks, H.~Mecking, Physics and phenomenology of strain hardening: the {FCC}
  case, Progress in Materials Science 48~(3) (2003) 171--273.
\newblock \href {https://doi.org/10.1016/s0079-6425(02)00003-8}
  {\path{doi:10.1016/s0079-6425(02)00003-8}}.

\bibitem{Taylor1934}
G.~I. Taylor, The mechanism of plastic deformation of crystals. part
  {II}.{\textemdash}comparison with observations, Proceedings of the Royal
  Society of London. Series A, Containing Papers of a Mathematical and Physical
  Character 145~(855) (1934) 388--404.
\newblock \href {https://doi.org/10.1098/rspa.1934.0107}
  {\path{doi:10.1098/rspa.1934.0107}}.

\bibitem{Taylor1934a}
G.~I. Taylor, The mechanism of plastic deformation of crystals. part
  i.{\textemdash}theoretical, Proceedings of the Royal Society of London.
  Series A, Containing Papers of a Mathematical and Physical Character
  145~(855) (1934) 362--387.
\newblock \href {https://doi.org/10.1098/rspa.1934.0106}
  {\path{doi:10.1098/rspa.1934.0106}}.

\bibitem{Franciosi1980}
P.~Franciosi, M.~Berveiller, A.~Zaoui, Latent hardening in copper and aluminium
  single crystals, Acta Metallurgica 28~(3) (1980) 273--283.
\newblock \href {https://doi.org/10.1016/0001-6160(80)90162-5}
  {\path{doi:10.1016/0001-6160(80)90162-5}}.

\bibitem{ma_dislocation_2006}
A.~Ma, F.~Roters, D.~Raabe,
  \href{http://www.sciencedirect.com/science/article/pii/S1359645406000747}{A
  dislocation density based constitutive model for crystal plasticity {FEM}
  including geometrically necessary dislocations}, Acta Materialia 54~(8)
  (2006) 2169--2179.
\newblock \href {https://doi.org/10.1016/j.actamat.2006.01.005}
  {\path{doi:10.1016/j.actamat.2006.01.005}}.
\newline\urlprefix\url{http://www.sciencedirect.com/science/article/pii/S1359645406000747}

\bibitem{Kubin2008}
L.~Kubin, B.~Devincre, T.~Hoc, Modeling dislocation storage rates and mean free
  paths in face-centered cubic crystals, Acta Materialia 56~(20) (2008)
  6040--6049.
\newblock \href {https://doi.org/10.1016/j.actamat.2008.08.012}
  {\path{doi:10.1016/j.actamat.2008.08.012}}.

\bibitem{alankar_explicit_2012}
A.~Alankar, D.~P. Field, H.~M. Zbib,
  \href{http://dx.doi.org/10.1080/14786435.2012.685964}{Explicit incorporation
  of cross-slip in a dislocation density-based crystal plasticity model},
  Philosophical Magazine 92~(24) (2012) 3084--3100.
\newblock \href {https://doi.org/10.1080/14786435.2012.685964}
  {\path{doi:10.1080/14786435.2012.685964}}.
\newline\urlprefix\url{http://dx.doi.org/10.1080/14786435.2012.685964}

\bibitem{Reuber.2014}
C.~Reuber, P.~Eisenlohr, F.~Roters, D.~Raabe, Dislocation density distribution
  around an indent in single-crystalline nickel: Comparing nonlocal crystal
  plasticity finite-element predictions with experiments, Acta materialia 71
  (2014) 333--348.

\bibitem{Franciosi_1982a}
P.~Franciosi, A.~Zaoui, Multislip in f.c.c. crystals a theoretical approach
  compared with experimental data, Acta Metallurgica 30~(8) (1982) 1627--1637.
\newblock \href {https://doi.org/10.1016/0001-6160(82)90184-5}
  {\path{doi:10.1016/0001-6160(82)90184-5}}.

\bibitem{Franciosi_1982b}
P.~Franciosi, A.~Zaoui, Multislip tests on copper crystals: A junctions
  hardening effect, Acta Metallurgica 30~(12) (1982) 2141--2151.
\newblock \href {https://doi.org/10.1016/0001-6160(82)90135-3}
  {\path{doi:10.1016/0001-6160(82)90135-3}}.

\bibitem{madec_dislocation_2002}
R.~Madec, B.~Devincre, L.~Kubin, From dislocation junctions to forest
  hardening, Physical review letters 89~(25) (2002) 255508.

\bibitem{Schoeck1972}
G.~Schoeck, R.~Frydman, The contribution of the dislocation forest to the flow
  stress, Physica Status Solidi (b) 53~(2) (1972) 661--673.
\newblock \href {https://doi.org/10.1002/pssb.2220530227}
  {\path{doi:10.1002/pssb.2220530227}}.

\bibitem{neuhaus_flow_1992}
R.~Neuhaus, C.~Schwink,
  \href{http://www.tandfonline.com/doi/abs/10.1080/01418619208205617}{On the
  flow stress of [100]- and [111]-oriented {Cu}-{Mn} single crystals: {A}
  transmission electron microscopy study}, Philosophical Magazine A 65~(6)
  (1992) 1463--1484.
\newblock \href {https://doi.org/10.1080/01418619208205617}
  {\path{doi:10.1080/01418619208205617}}.
\newline\urlprefix\url{http://www.tandfonline.com/doi/abs/10.1080/01418619208205617}

\bibitem{madec_role_2003}
R.~Madec, B.~Devincre, L.~Kubin, T.~Hoc, D.~Rodney,
  \href{http://science.sciencemag.org/content/301/5641/1879}{The {Role} of
  {Collinear} {Interaction} in {Dislocation}-{Induced} {Hardening}}, Science
  301~(5641) (2003) 1879--1882.
\newblock \href {https://doi.org/10.1126/science.1085477}
  {\path{doi:10.1126/science.1085477}}.
\newline\urlprefix\url{http://science.sciencemag.org/content/301/5641/1879}

\bibitem{Devincre2006}
B.~Devincre, L.~Kubin, T.~Hoc, Physical analyses of crystal plasticity by {DD}
  simulations, Scripta Materialia 54~(5) (2006) 741--746.
\newblock \href {https://doi.org/10.1016/j.scriptamat.2005.10.066}
  {\path{doi:10.1016/j.scriptamat.2005.10.066}}.

\bibitem{Madec2008}
R.~Madec, L.~P. Kubin, Second-order junctions and strain hardening in bcc and
  fcc crystals, Scripta Materialia 58~(9) (2008) 767--770.
\newblock \href {https://doi.org/10.1016/j.scriptamat.2007.12.032}
  {\path{doi:10.1016/j.scriptamat.2007.12.032}}.

\bibitem{Arsenlis.2002}
A.~Arsenlis, D.~M. Parks, Modeling the evolution of crystallographic
  dislocation density in crystal plasticity, Journal of the Mechanics and
  Physics of Solids 50~(9) (2002) 1979--2009.
\newblock \href {https://doi.org/10.1016/S0022-5096(01)00134-X}
  {\path{doi:10.1016/S0022-5096(01)00134-X}}.

\bibitem{leung.2015}
H.~Leung, P.~Leung, B.~Cheng, A.~Ngan,
  \href{http://www.sciencedirect.com/science/article/pii/S0749641914001934}{A
  new dislocation-density-function dynamics scheme for computational crystal
  plasticity by explicit consideration of dislocation elastic interactions},
  International Journal of Plasticity 67 (2015) 1--25.
\newblock \href {https://doi.org/10.1016/j.ijplas.2014.09.009}
  {\path{doi:10.1016/j.ijplas.2014.09.009}}.
\newline\urlprefix\url{http://www.sciencedirect.com/science/article/pii/S0749641914001934}

\bibitem{Hochrainer2007}
T.~Hochrainer, Evolving systems of curved dislocations: mathematical
  foundations of a statistical theory, Ph.D. thesis, Karlsruher Institut fuer
  Technologie (KIT) (2007).
\newblock \href {https://doi.org/10.13140/RG.2.1.1630.6407}
  {\path{doi:10.13140/RG.2.1.1630.6407}}.

\bibitem{schulz_mesoscale_2019}
K.~Schulz, L.~Wagner, C.~Wieners, A mesoscale continuum approach of dislocation
  dynamics and the approximation by a runge-kutta discontinuous galerkin
  method, International Journal of Plasticity 120 (2019) 248 -- 261.
\newblock \href {https://doi.org/https://doi.org/10.1016/j.ijplas.2019.05.003}
  {\path{doi:https://doi.org/10.1016/j.ijplas.2019.05.003}}.

\bibitem{schulz_dislocation-density_2017}
K.~Schulz, M.~Sudmanns, P.~Gumbsch, Dislocation-density based description of
  the deformation of a composite material, Modelling and Simulation in
  Materials Science and Engineering 25~(6) (2017) 064003.

\bibitem{Ma2004}
A.~Ma, F.~Roters, A constitutive model for fcc single crystals based on
  dislocation densities and its application to uniaxial compression of
  aluminium single crystals, Acta Materialia 52~(12) (2004) 3603--3612.
\newblock \href {https://doi.org/10.1016/j.actamat.2004.04.012}
  {\path{doi:10.1016/j.actamat.2004.04.012}}.

\bibitem{li_predicting_2014}
D.~Li, H.~Zbib, X.~Sun, M.~Khaleel,
  \href{http://www.sciencedirect.com/science/article/pii/S0749641913000259}{Predicting
  plastic flow and irradiation hardening of iron single crystal with
  mechanism-based continuum dislocation dynamics}, International Journal of
  Plasticity 52 (2014) 3--17.
\newblock \href {https://doi.org/10.1016/j.ijplas.2013.01.015}
  {\path{doi:10.1016/j.ijplas.2013.01.015}}.
\newline\urlprefix\url{http://www.sciencedirect.com/science/article/pii/S0749641913000259}

\bibitem{Sudmanns_2020}
M.~Sudmanns, J.~Bach, D.~Weygand, K.~Schulz,
  \href{https://doi.org/10.1088/1361-651x/ab97ef}{Data-driven exploration and
  continuum modeling of dislocation networks}, Modelling and Simulation in
  Materials Science and Engineering 28~(6) (2020) 065001.
\newblock \href {https://doi.org/10.1088/1361-651x/ab97ef}
  {\path{doi:10.1088/1361-651x/ab97ef}}.
\newline\urlprefix\url{https://doi.org/10.1088/1361-651x/ab97ef}

\bibitem{monavari_annihilation_2018}
M.~Monavari, M.~Zaiser,
  \href{https://materialstheory.springeropen.com/articles/10.1186/s41313-018-0010-z}{Annihilation
  and sources in continuum dislocation dynamics}, Materials Theory 2~(1)
  (2018).
\newblock \href {https://doi.org/10.1186/s41313-018-0010-z}
  {\path{doi:10.1186/s41313-018-0010-z}}.
\newline\urlprefix\url{https://materialstheory.springeropen.com/articles/10.1186/s41313-018-0010-z}

\bibitem{sudmanns_dislocation_2019}
M.~Sudmanns, M.~Stricker, D.~Weygand, T.~Hochrainer, K.~Schulz,
  \href{http://www.sciencedirect.com/science/article/pii/S0022509619306428}{Dislocation
  multiplication by cross-slip and glissile reaction in a dislocation based
  continuum formulation of crystal plasticity}, Journal of the Mechanics and
  Physics of Solids 132 (2019) 103695.
\newblock \href {https://doi.org/10.1016/j.jmps.2019.103695}
  {\path{doi:10.1016/j.jmps.2019.103695}}.
\newline\urlprefix\url{http://www.sciencedirect.com/science/article/pii/S0022509619306428}

\bibitem{Giessen2020}
E.~van~der Giessen, P.~A. Schultz, N.~Bertin, V.~V. Bulatov, W.~Cai,
  G.~Cs{\'{a}}nyi, S.~M. Foiles, M.~G.~D. Geers, C.~Gonz{\'{a}}lez, M.~Hütter,
  W.~K. Kim, D.~M. Kochmann, J.~LLorca, A.~E. Mattsson, J.~Rottler, A.~Shluger,
  R.~B. Sills, I.~Steinbach, A.~Strachan, E.~B. Tadmor, Roadmap on multiscale
  materials modeling, Modelling and Simulation in Materials Science and
  Engineering 28~(4) (2020) 043001.
\newblock \href {https://doi.org/10.1088/1361-651x/ab7150}
  {\path{doi:10.1088/1361-651x/ab7150}}.

\bibitem{Schulz2018}
K.~Schulz, S.~Schmitt, Discrete-continuum transition: A discussion of the
  continuum limit, Technische Mechanik; 38; 1; 126-134; ISSN 2199-9244 (2018).
\newblock \href {https://doi.org/10.24352/UB.OVGU-2018-012}
  {\path{doi:10.24352/UB.OVGU-2018-012}}.

\bibitem{Morgan2020}
D.~Morgan, R.~Jacobs, Opportunities and challenges for machine learning in
  materials science, Annual Review of Materials Research 50~(1) (2020) 71--103.
\newblock \href {https://doi.org/10.1146/annurev-matsci-070218-010015}
  {\path{doi:10.1146/annurev-matsci-070218-010015}}.

\bibitem{Huber_2020}
N.~Huber, S.~R. Kalidindi, B.~Klusemann, C.~J. Cyron,
  \href{https://www.frontiersin.org/article/10.3389/fmats.2020.00051}{Editorial:
  Machine learning and data mining in materials science}, Frontiers in
  Materials 7 (2020) 51.
\newblock \href {https://doi.org/10.3389/fmats.2020.00051}
  {\path{doi:10.3389/fmats.2020.00051}}.
\newline\urlprefix\url{https://www.frontiersin.org/article/10.3389/fmats.2020.00051}

\bibitem{Steinberger2019}
D.~Steinberger, H.~Song, S.~Sandfeld, Machine learning-based classification of
  dislocation microstructures, Frontiers in Materials 6 (jun 2019).
\newblock \href {https://doi.org/10.3389/fmats.2019.00141}
  {\path{doi:10.3389/fmats.2019.00141}}.

\bibitem{Salmenjoki2018}
H.~Salmenjoki, M.~J. Alava, L.~Laurson, Machine learning plastic deformation of
  crystals, Nature Communications 9~(1) (dec 2018).
\newblock \href {https://doi.org/10.1038/s41467-018-07737-2}
  {\path{doi:10.1038/s41467-018-07737-2}}.

\bibitem{Mangal2018}
A.~Mangal, E.~A. Holm, Applied machine learning to predict stress hotspots i:
  Face centered cubic materials, International Journal of Plasticity 111 (2018)
  122--134.
\newblock \href {https://doi.org/10.1016/j.ijplas.2018.07.013}
  {\path{doi:10.1016/j.ijplas.2018.07.013}}.

\bibitem{Akhondzadeh2021}
S.~Akhondzadeh, N.~Bertin, R.~B. Sills, W.~Cai, Slip-free multiplication and
  complexity of dislocation networks in {FCC} metals, Materials Theory 5~(1)
  (mar 2021).
\newblock \href {https://doi.org/10.1186/s41313-020-00024-y}
  {\path{doi:10.1186/s41313-020-00024-y}}.

\bibitem{Akhondzadeh2020}
S.~Akhondzadeh, R.~B. Sills, N.~Bertin, W.~Cai, Dislocation density-based
  plasticity model from massive discrete dislocation dynamics database, Journal
  of the Mechanics and Physics of Solids 145 (2020) 104152.
\newblock \href {https://doi.org/10.1016/j.jmps.2020.104152}
  {\path{doi:10.1016/j.jmps.2020.104152}}.

\bibitem{sills_dislocation_2018}
R.~B. Sills, N.~Bertin, A.~Aghaei, W.~Cai, Dislocation {Networks} and the
  {Microstructural} {Origin} of {Strain} {Hardening}, Physical Review Letters
  121~(8) (2018) 085501.
\newblock \href {https://doi.org/10.1103/PhysRevLett.121.085501}
  {\path{doi:10.1103/PhysRevLett.121.085501}}.

\bibitem{Davoudi2018}
K.~M. Davoudi, J.~J. Vlassak, Dislocation evolution during plastic deformation:
  Equations vs. discrete dislocation dynamics study, Journal of Applied Physics
  123~(8) (2018) 085302.
\newblock \href {https://doi.org/10.1063/1.5013213}
  {\path{doi:10.1063/1.5013213}}.

\bibitem{Stricker2015}
M.~Stricker, D.~Weygand, Dislocation multiplication mechanisms {\textendash}
  glissile junctions and their role on the plastic deformation at the
  microscale, Acta Materialia 99 (2015) 130--139.
\newblock \href {https://doi.org/10.1016/j.actamat.2015.07.073}
  {\path{doi:10.1016/j.actamat.2015.07.073}}.

\bibitem{Weygand_2002}
D.~Weygand, L.~H. Friedman, E.~V. der Giessen, A.~Needleman,
  \href{https://doi.org/10.1088/0965-0393/10/4/306}{Aspects of boundary-value
  problem solutions with three-dimensional dislocation dynamics}, Modelling and
  Simulation in Materials Science and Engineering 10~(4) (2002) 437--468.
\newblock \href {https://doi.org/10.1088/0965-0393/10/4/306}
  {\path{doi:10.1088/0965-0393/10/4/306}}.
\newline\urlprefix\url{https://doi.org/10.1088/0965-0393/10/4/306}

\bibitem{ElAchkar2019}
T.~El-Achkar, D.~Weygand, Analysis of dislocation microstructure
  characteristics of surface grains under cyclic loading by discrete
  dislocation dynamics, Modelling and Simulation in Materials Science and
  Engineering 27~(5) (2019) 055004.
\newblock \href {https://doi.org/10.1088/1361-651x/ab1b7c}
  {\path{doi:10.1088/1361-651x/ab1b7c}}.

\bibitem{Forsyth2019}
D.~Forsyth, Applied Machine Learning, Springer International Publishing, 2019.
\newblock \href {https://doi.org/10.1007/978-3-030-18114-7}
  {\path{doi:10.1007/978-3-030-18114-7}}.

\bibitem{James2013}
G.~James, D.~Witten, T.~Hastie, R.~Tibshirani,
  \href{https://doi.org/10.1007/978-1-4614-7138-7}{An Introduction to
  Statistical Learning}, Springer New York, 2013.
\newblock \href {https://doi.org/10.1007/978-1-4614-7138-7}
  {\path{doi:10.1007/978-1-4614-7138-7}}.
\newline\urlprefix\url{https://doi.org/10.1007/978-1-4614-7138-7}

\bibitem{Unpingco2019}
J.~Unpingco, Python for Probability, Statistics, and Machine Learning, Springer
  International Publishing, 2019.
\newblock \href {https://doi.org/10.1007/978-3-030-18545-9}
  {\path{doi:10.1007/978-3-030-18545-9}}.

\bibitem{Roters2010}
F.~Roters, P.~Eisenlohr, L.~Hantcherli, D.~Tjahjanto, T.~Bieler, D.~Raabe,
  Overview of constitutive laws, kinematics, homogenization and multiscale
  methods in crystal plasticity finite-element modeling: Theory, experiments,
  applications, Acta Materialia 58~(4) (2010) 1152--1211.
\newblock \href {https://doi.org/10.1016/j.actamat.2009.10.058}
  {\path{doi:10.1016/j.actamat.2009.10.058}}.

\bibitem{Orowan1934a}
E.~Orowan, Zur kristallplastizität. i, Zeitschrift für Physik,
  89(9-10):605–613 (1934).

\bibitem{Orowan1934b}
E.~Orowan, Zur kristallplastizität. ii, Zeitschrift für Physik,
  89(9-10):614–633 (1934).

\bibitem{Orowan1934c}
E.~Orowan, Zur kristallplastizität. iii, Zeitschrift für Physik,
  89(9-10):634-659 (1934).

\bibitem{Zoller2021}
K.~Zoller, S.~Kal{\'{a}}cska, P.~D. Isp{\'{a}}novity, K.~Schulz, Microstructure
  evolution of compressed micropillars investigated by in situ {HR}-{EBSD}
  analysis and dislocation density simulations, Comptes Rendus. Physique
  22~(S3) (2021) 1--27.
\newblock \href {https://doi.org/10.5802/crphys.55}
  {\path{doi:10.5802/crphys.55}}.

\bibitem{Weygand_2001}
D.~Weygand, L.~Friedman, E.~{van der Giessen}, A.~Needleman,
  \href{http://www.sciencedirect.com/science/article/pii/S0921509300016324}{Discrete
  dislocation modeling in three-dimensional confined volumes}, Materials
  Science and Engineering: A 309-310 (2001) 420 -- 424, dislocations 2000: An
  International Conference on the Fundamentals of Plastic Deformation.
\newblock \href {https://doi.org/https://doi.org/10.1016/S0921-5093(00)01632-4}
  {\path{doi:https://doi.org/10.1016/S0921-5093(00)01632-4}}.
\newline\urlprefix\url{http://www.sciencedirect.com/science/article/pii/S0921509300016324}

\bibitem{Weygand2005}
D.~Weygand, P.~Gumbsch, Study of dislocation reactions and rearrangements under
  different loading conditions, Materials Science and Engineering: A 400-401
  (2005) 158--161.
\newblock \href {https://doi.org/10.1016/j.msea.2005.03.102}
  {\path{doi:10.1016/j.msea.2005.03.102}}.

\bibitem{Schmid1935}
E.~Schmid, W.~Boas, Kristallplastizität, Springer Berlin Heidelberg, 1935.
\newblock \href {https://doi.org/10.1007/978-3-662-34532-0}
  {\path{doi:10.1007/978-3-662-34532-0}}.

\bibitem{Motz2009}
C.~Motz, D.~Weygand, J.~Senger, P.~Gumbsch, Initial dislocation structures in
  3-d discrete dislocation dynamics and their influence on microscale
  plasticity, Acta Materialia 57~(6) (2009) 1744--1754.
\newblock \href {https://doi.org/10.1016/j.actamat.2008.12.020}
  {\path{doi:10.1016/j.actamat.2008.12.020}}.

\bibitem{Stricker_2018}
M.~Stricker, M.~Sudmanns, K.~Schulz, T.~Hochrainer, D.~Weygand,
  \href{http://www.sciencedirect.com/science/article/pii/S0022509618301157}{Dislocation
  multiplication in stage ii deformation of fcc multi-slip single crystals},
  Journal of the Mechanics and Physics of Solids 119 (2018) 319 -- 333.
\newblock \href {https://doi.org/https://doi.org/10.1016/j.jmps.2018.07.003}
  {\path{doi:https://doi.org/10.1016/j.jmps.2018.07.003}}.
\newline\urlprefix\url{http://www.sciencedirect.com/science/article/pii/S0022509618301157}

\bibitem{Kraft2010}
O.~Kraft, P.~P.~A. Gruber, R.~M{\"{o}}nig, D.~Weygand,
  \href{http://www.annualreviews.org/doi/abs/10.1146/annurev-matsci-082908-145409}{{Plasticity
  in Confined Dimensions}}, Annual Review of Materials Research 40~(1) (2010)
  293--317.
\newblock \href {https://doi.org/10.1146/annurev-matsci-082908-145409}
  {\path{doi:10.1146/annurev-matsci-082908-145409}}.
\newline\urlprefix\url{http://www.annualreviews.org/doi/abs/10.1146/annurev-matsci-082908-145409}

\bibitem{el-awady_unravelling_2015}
J.~A. El-Awady, \href{http://www.nature.com/articles/ncomms6926}{Unravelling
  the physics of size-dependent dislocation-mediated plasticity}, Nature
  Communications 6~(1) (Dec. 2015).
\newblock \href {https://doi.org/10.1038/ncomms6926}
  {\path{doi:10.1038/ncomms6926}}.
\newline\urlprefix\url{http://www.nature.com/articles/ncomms6926}

\bibitem{Bertin2020}
N.~Bertin, R.~B. Sills, W.~Cai,
  \href{https://doi.org/10.1146/annurev-matsci-091819-015500}{Frontiers in the
  simulation of dislocations}, Annual Review of Materials Research 50~(1)
  (2020) 437--464.
\newblock \href
  {http://arxiv.org/abs/https://doi.org/10.1146/annurev-matsci-091819-015500}
  {\path{arXiv:https://doi.org/10.1146/annurev-matsci-091819-015500}}, \href
  {https://doi.org/10.1146/annurev-matsci-091819-015500}
  {\path{doi:10.1146/annurev-matsci-091819-015500}}.
\newline\urlprefix\url{https://doi.org/10.1146/annurev-matsci-091819-015500}

\bibitem{Luecke1952}
K.~Lücke, H.~Lange, Über die form der verfestigungskurve von
  reinstaluminiumkristallen und die bildung von deformationsbändern,
  International Journal of Materials Research 43~(2) (1952) 55--66.
\newblock \href {https://doi.org/10.1515/ijmr-1952-430209}
  {\path{doi:10.1515/ijmr-1952-430209}}.

\bibitem{Hosford_1960}
W.~Hosford, R.~Fleischer, W.~Backofen, Tensile deformation of aluminum single
  crystals at low temperatures, Acta Metallurgica 8~(3) (1960) 187--199.
\newblock \href {https://doi.org/10.1016/0001-6160(60)90127-9}
  {\path{doi:10.1016/0001-6160(60)90127-9}}.

\bibitem{takeuchi_work_1975}
T.~Takeuchi, Work hardening of copper single crystals with multiple glide
  orientations, Transactions of the Japan Institute of Metals 16~(10) (1975)
  629--640.

\bibitem{Weygand2014}
D.~Weygand,
  \href{http://journals.cambridge.org/abstract_S1946427414003625}{{Mechanics
  and Dislocation Structures at the Micro-Scale: Insights on Dislocation
  Multiplication Mechanisms from Discrete Dislocation Dynamics Simulations}},
  MRS Proceedings 1651 (2014) mrsf13--1651--kk07--02.
\newblock \href {https://doi.org/10.1557/opl.2014.362}
  {\path{doi:10.1557/opl.2014.362}}.
\newline\urlprefix\url{http://journals.cambridge.org/abstract_S1946427414003625}

\bibitem{Ma2006}
A.~Ma, F.~Roters, D.~Raabe, Studying the effect of grain boundaries in
  dislocation density based crystal-plasticity finite element simulations,
  International Journal of Solids and Structures 43~(24) (2006) 7287--7303.
\newblock \href {https://doi.org/10.1016/j.ijsolstr.2006.07.006}
  {\path{doi:10.1016/j.ijsolstr.2006.07.006}}.

\bibitem{Fan2021}
H.~Fan, Q.~Wang, J.~A. El-Awady, D.~Raabe, M.~Zaiser, Strain rate dependency of
  dislocation plasticity, Nature Communications 12~(1) (mar 2021).
\newblock \href {https://doi.org/10.1038/s41467-021-21939-1}
  {\path{doi:10.1038/s41467-021-21939-1}}.

\bibitem{hussein_microstructurally_2015}
A.~M. Hussein, S.~I. Rao, M.~D. Uchic, D.~M. Dimiduk, J.~A. El-Awady,
  \href{http://www.sciencedirect.com/science/article/pii/S1359645414008465}{Microstructurally
  based cross-slip mechanisms and their effects on dislocation microstructure
  evolution in fcc crystals}, Acta Materialia 85 (2015) 180--190.
\newblock \href {https://doi.org/10.1016/j.actamat.2014.10.067}
  {\path{doi:10.1016/j.actamat.2014.10.067}}.
\newline\urlprefix\url{http://www.sciencedirect.com/science/article/pii/S1359645414008465}

\bibitem{alankar_determination_2012}
A.~Alankar, I.~N. Mastorakos, D.~P. Field, H.~M. Zbib,
  \href{http://dx.doi.org/10.1115/1.4005917}{Determination of {Dislocation}
  {Interaction} {Strengths} {Using} {Discrete} {Dislocation} {Dynamics} of
  {Curved} {Dislocations}}, Journal of Engineering Materials and Technology
  134~(2) (2012) 021018--021018--4.
\newblock \href {https://doi.org/10.1115/1.4005917}
  {\path{doi:10.1115/1.4005917}}.
\newline\urlprefix\url{http://dx.doi.org/10.1115/1.4005917}

\end{thebibliography}

%% \bibitem[Author(year)]{label}
%% Text of bibliographic item

% \bibitem[ ()]{}

% \end{thebibliography}
\end{document}